\documentclass[preprint,12pt]{elsarticle}

\usepackage{amssymb}
\usepackage{amsmath} 
\usepackage{amsfonts} 
\usepackage{graphicx} 
\usepackage{listings}
\usepackage{xcolor}
\usepackage{tcolorbox}
\usepackage{caption}
\usepackage{subcaption}
\usepackage{hyperref}
\usepackage{bm}
\usepackage[T1]{fontenc}
\usepackage{multirow}
\usepackage{array}
\usepackage{comment}

\lstdefinestyle{mystyle}{
  backgroundcolor=\color{lightgray},   
  keywordstyle=\color{blue},
  identifierstyle=\color{black},
  basicstyle=\ttfamily,
  breaklines=true,                 
  captionpos=b,                    
  keepspaces=true,                 
  numbers=none,                    
  showspaces=false,                
  showstringspaces=false,
  showtabs=false,                  
  tabsize=2
}

\journal{Results in Engineering}

\begin{document}

\begin{frontmatter}



\title{Beyond Language: Applying MLX Transformers to Engineering Physics}


\author[inst1]{Stavros Kassinos\corref{cor1}}
\cortext[cor1]{Corresponding author}
\ead{kassinos@ucy.ac.cy}

\affiliation[inst1]{organization={Computational Sciences Laboratory, Department of Mechanical Engineering, University of Cyprus},
            addressline={1 University Avenue}, 
            city={Aglantzia},
            postcode={2109}, 
            state={Nicosia},
            country={Cyprus}
            }

\author[inst2]{Alessio Alexiadis}
\ead{a.alexiadis@bham.ac.uk}

\affiliation[inst2]{organization={School of Chemical Engineering, University of Birmingham},
            addressline={Edgbaston}, 
            city={Birmingham, B15 2TT},
            postcode={22222}, 
            state={State Two},
            country={United Kingdom}}

\begin{abstract}
Transformer Neural Networks are driving an explosion of activity and discovery in the field of Large Language Models (LLMs). In contrast, there have been only a few attempts to apply Transformers  in engineering physics.  Aiming to offer an easy entry point to physics-centric Transformers,  we introduce a physics-informed Transformer  model for solving the heat conduction problem in a 2D plate with Dirichlet boundary conditions. The model is implemented in the machine learning framework MLX and leverages the unified memory of Apple M-series processors. The use of MLX means that the models can be trained and perform predictions efficiently on personal machines with only modest memory requirements. To train, validate and test the Transformer  model we solve the 2D heat conduction problem using central finite differences. Each finite difference solution in these sets is initialized with four random Dirichlet boundary conditions, a uniform but random internal temperature distribution and a randomly selected thermal diffusivity. Validation is performed in-line during training to monitor against over-fitting. The excellent performance of the trained model is demonstrated by predicting the evolution of the temperature field to steady state for the unseen test set of conditions.
\end{abstract}

\begin{graphicalabstract}
\includegraphics[width=0.5\textwidth]{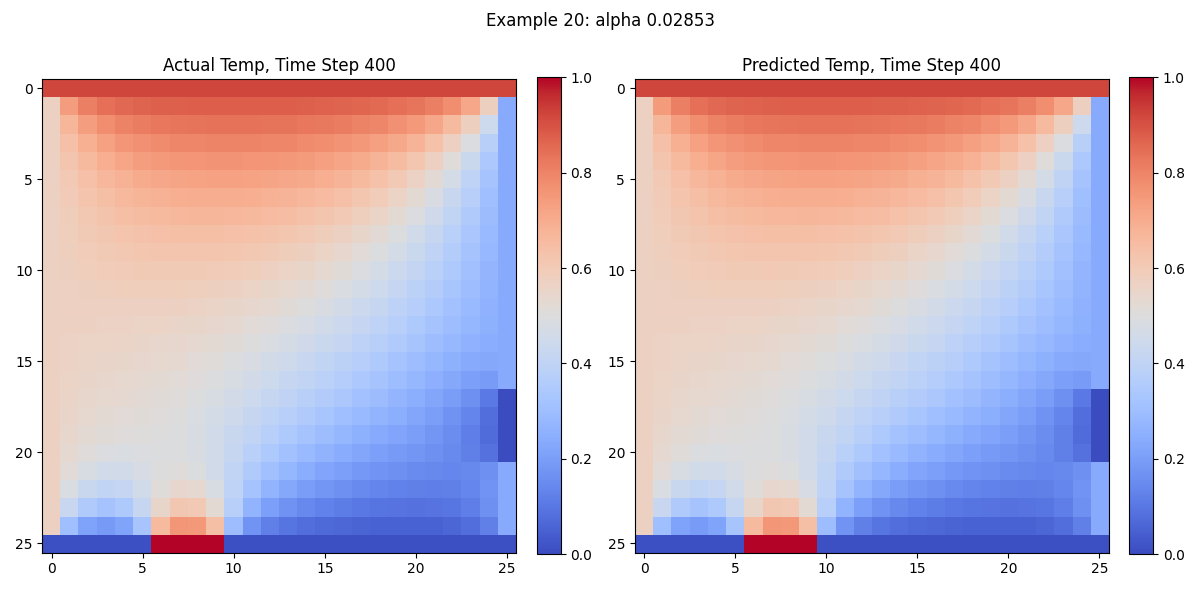}

\end{graphicalabstract}

\begin{highlights}
\item Parallels are drawn between key features of Transformers and established concepts in mathematical physics.
\item A standard Transformer architecture is trained using the MLX framework.
\item MLX is optimized for the unified memory architecture of Apple M-series processors and the code can run on personal machines.
\item The trained physics-informed Transformer achieves excellent inference performance.
\end{highlights}


\end{frontmatter}

\section{Introduction and motivation}
\label{sec:Introduction}
The landmark paper "Attention is all you need" introduced Transformers  in 2017 \cite{attention_is_all_you_need}. Since then, Transformers  have expanded their reach into a wide range of different applications within the field of Artificial Intelligence (AI). To date, however, the potential of Transformer-based neural networks (NNs) in physics and engineering remains relatively unexplored. One reason for this is that the community is faced with the challenge of translating the available information on Transformers, which is mostly framed in the context of Natural Language Processing (NLP), to a completely different paradigm.

A few recent studies have introduced Transformers  in conjunction with physical problems \cite{Yousif2023-li}, \cite{Zhao2023-zq}, \cite{Lorsung2024-vh}, \cite{Li2022-xe}, \cite{Alexiadis_Ghiassi2024}. These works employ specialized, ad-hoc architectures that are tailored to applying physics-informed machine  learning (ML) to specific challenges expressed as Partial Differential Equations (PDEs). They have demonstrated the potential of Transformers  in solving complex physical problems by incorporating specialized architectures and mechanisms designed to handle specific PDEs.

In contrast, this work focuses on demonstrating the general applicability and efficiency of Transformers.  It highlights the potential of using `pure' Transformer  models, not as part of an ad-hoc physics-informed architecture, in engineering physics. This is demonstrated in conjunction with the practical benefits of Transformers’ parallel processing capabilities, especially when combined with efficient frameworks like MLX \cite{MLX_GitHub}. 

MLX entered the AI and ML scene in the Fall of 2023 and has quickly gained attention for its lean and highly efficient coding platform. It leverages the unified memory hardware of Apple silicon (currently M1, M2, and M3 processors, with M4 processors coming soon). This integration means that tensors do not need to be explicitly moved between the CPU and GPU, keeping the code lean and avoiding unnecessary complexity. The synergy of unified memory with the `lazy execution' paradigm used in MLX enables efficient computations on personal computers (see~\ref{sec:appendixA}).

This paper, therefore, addresses the research question of whether standard Transformer  architectures are capable of `learning' differential operators by themselves, without the support of ad-hoc solutions. To provide a friendly introduction to these physics-centric Transformers,  we train a Transformer  model to predict the solutions of heat conduction in a 2D plate with Dirichlet boundary conditions (BCs). This particular setup strikes a good balance between being simple enough to avoid obscuring basic concepts and being sufficiently rich in long-term temporal and spatial relationships. It has been chosen to provide an opportunity to explore the most important features of Transformers.  In fact, Transformers  are specifically designed to discover and model long-term patterns in data due to their self-attention mechanism. In our case study, we will test their ability to capture the long-term evolution of temperature profiles resulting from different sets of BCs. This is in contrast with particle methods such as Molecular Dynamics, which are modeled by large systems of Ordinary Differential Equations (ODEs) rather than PDEs. In these systems, the positions and velocities of particles quickly become decorrelated from their initial values due to the inherently chaotic nature of the dynamics. Consequently, Transformers  have not shown significant advantages in these contexts \cite{Alexiadis2024}, and often other approaches might be preferred \cite{Alexiadis2023}.

\section{A High-Level Introduction to Transformers}
\label{sec:sub:IntroTransformers}

The goal of this work is to provide an introduction to Transformers  for engineering applications when the data are numerical, such as those coming from time series, spatial data, or, as in the case under study, solutions of differential equations. However, to begin with, let us start with a very high-level view of how Transformers  work for their intended scope, NLP.

Introductions to Transformer models typically fall into two categories: they are either highly simplified to support those with minimal machine learning knowledge, or they are highly detailed, targeting those with a strong background in computer science. In this article, we aim to strike a balance by targeting an intermediate audience—those who find the first approach too basic and the second too complex.

We assume our readers are engineers or scientists with a solid understanding of mathematical concepts like linear algebra, approximation theory, signal processing, and integral transforms, such as Fourier analysis. Additionally, we expect that our readers have a basic familiarity with artificial neural networks, backpropagation, stochastic gradient descent, and related topics. This profile aligns with many scientists and engineers who have recently begun exploring the basics of machine learning but do not yet consider themselves experts.

According to well-established educational theories ~\cite{Brown2014} , new concepts are more effectively learned when connected to prior knowledge. Therefore, we will build on our readers' existing knowledge to gradually introduce the Transformer architecture. To gain some intuition on the subject, we will show that many concepts, which may initially seem complex or unfamiliar, are conceptually similar to ideas the reader already understands. For instance, we will explore how the self-attention mechanism in Transformers parallels the Fourier transform, as both employ a similar mathematical framework to uncover patterns and relationships across data points—whether these are sequences of words or frequencies in a signal. Additionally, we will draw comparisons between positional embeddings and the time-localization capabilities of wavelet transforms, which introduce temporal information that the Fourier transform alone does not capture. Furthermore, we’ll extend this analogy by comparing cross-attention in Transformers to cross-correlation in signal processing. While cross-correlation measures the similarity between two signals as a function of the time-lag applied to one of them, cross-attention in Transformers similarly aligns and integrates information from two different sequences (the encoder's and decoder's outputs) to enhance the model’s understanding and prediction capability. Therefore, readers familiar with concepts like Fourier analysis, wavelet transforms, or cross-correlation can relate these to the ideas such as self-attention and cross-attention in Transformers. These connections help demystify what might initially seem like exotic concepts by anchoring them to existing knowledge.\\

We will begin our exploration by initially looking at the Transformer as a black box during inference.

\begin{figure}[h!!]
    \centering
    \includegraphics[width=0.39\textwidth]{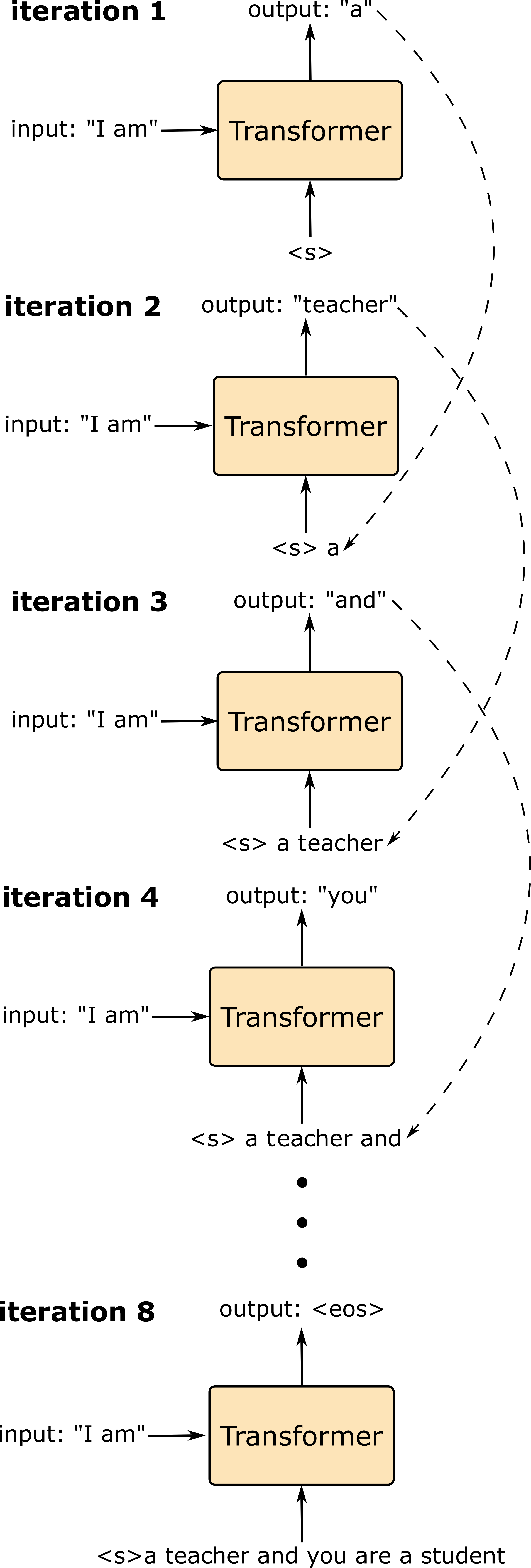}
    \caption{Step-by-step process of text generation using a Transformer model. At each iteration, the model takes the initial input ("I am") and the previously generated output to predict the next token. The model adds a special token \texttt{<s>} to the user's input, indicating the start of the sequence. The process continues, adding tokens to the sequence, until the end-of-sequence token \texttt{<eos>} is generated.}
    \label{fig:iterations}
\end{figure}
\break

Figure~\ref{fig:iterations} shows the iterative process of a Transformer  model generating a sentence, starting with an initial input and progressively adding words to construct a complete sentence. For now, we will focus only on inference and not training. Let us see how it works through each iteration, assuming that the initial user's input are the words "I am":
\begin{itemize}
    \item \textbf{Iteration 1:}
    \begin{itemize}
        \item \textbf{User Input:} "I am"\\
        This is the initial fixed input provided to the Transformer .
        \item \textbf{Transformer Input:} "$<$s$>$"\\
        The Transformer  starts by using the start-of-sequence token "$<$s$>$" as input.
        \item \textbf{Output:} "a"\\
        The Transformer  processes the user input "I am" along with the start-of-sequence token "$<$s$>$" and predicts "a" as the most probable next word.
    \end{itemize}
    \item \textbf{Iteration 2:}
    \begin{itemize}
        \item \textbf{User Input:} "I am"\\
        The user input remains the same for the whole inference process.
        \item \textbf{Transformer Input:} "$<$s$>$ a"\\
        Together with the user input "I am", the Transformer  now uses the start-of-sequence token followed by the previously generated word "a" as input.
        \item \textbf{Output:} "teacher"\\
        Using the input sequences "I am" and "$<$s$>$ a", the Transformer  predicts the next word "teacher".
    \end{itemize}
    \item \textbf{Iteration 3:}
    \begin{itemize}
        \item \textbf{User Input:} "I am"
        \item \textbf{Transformer Input:} "$<$s$>$ a teacher"\\
        The Transformer  uses the user input "I am" together with the sequence "$<$s$>$ a teacher" as input.
        \item \textbf{Output:} "and"\\
        Based on these inputs, the Transformer  predicts the next word "and".
    \end{itemize}
\end{itemize}

And the process continues until the Transformer  predicts the end-of-sequence token "$<$eos$>$", indicating that the sentence is complete. Thus, one word at a time, the Transformer  completes the user input to form the sentence "I am a teacher and you a student".

This very high-level overview requires some clarifications. First, the model does not always produce the same word for a given input. Instead, it generates a set of probabilities for each potential next word in the sequence and then chooses one word from this set based on these probabilities. Second, the model does not process words directly but uses numerical representations. Each word is converted into a numerical vector, called an embedding, before being processed by the Transformer. Therefore, the Transformer  architecture, although originally developed for NLP, can be applied to any sequence of numerical vectors. These vectors can represent not only words but also time series, spatial data, or solutions of differential equations. We leverage this flexibility of Transformers  for the heat conduction problem presented in Section~\ref{sec:Methods}.

\subsection{Understanding Transformers One Step Further: The Encoder and Decoder Architecture}

\begin{figure}[h!!]
    \centering
    \includegraphics[width=0.5\textwidth]{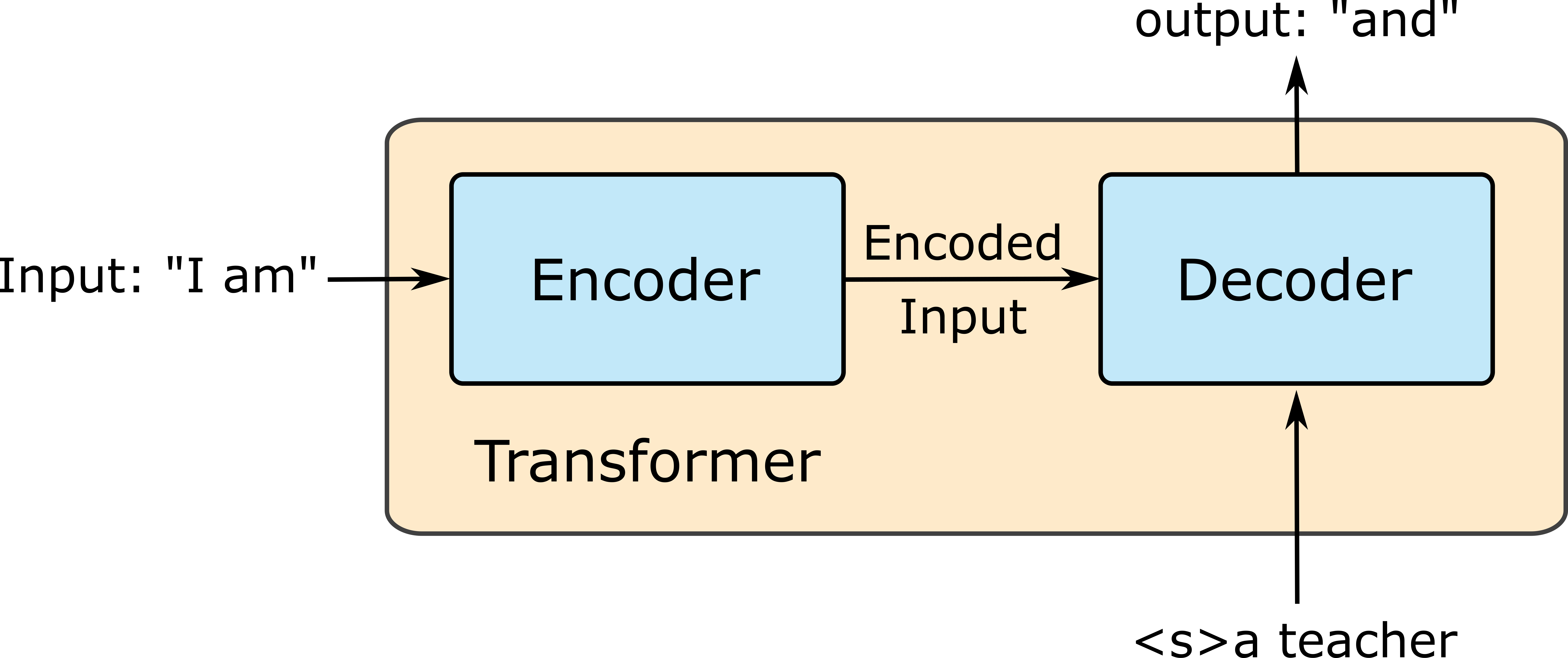}
    \caption{Encoder-Decoder architecture. This figure corresponds to Iteration 3 in Figure~\ref{fig:iterations}, where the encoder has processed the input ("I am") into an encoded representation. The decoder uses this encoded input and the previously generated output (\texttt{<s>} a teacher) to predict the next word in the sequence ("and").}
    \label{fig:encoder_decoder}
\end{figure}

In the previous section, we provided a very high-level introduction to Transformers,  focusing on how they generate text in natural language processing (NLP) tasks. Now, we will take a step further by introducing the core components of the Transformer  architecture: the encoder and decoder. For now, we will still consider these components as black boxes. The Transformer  has two components, the Encoder and the Decoder Figure~\ref{fig:encoder_decoder}.

\break

The encoder's primary function is to process the user input sequence and transform it into a set of abstract, high-dimensional representations. These representations capture the essential information about the input data, allowing the model to understand its context and meaning. To provide a engineering perspective on how Transformers  work, we can relate the underlying ideas to concepts in mathematics or physics that are familiar to engineers. In this case, we can think of the encoder as performing a transformation on the input sequence, converting it into a different representation of the same input.

In engineering, we often use transformations to analyze data. For example, the Fourier transform takes a time-domain signal and converts it into a frequency-domain representation, revealing the different frequencies present in the original signal. Similarly, the encoder in a Transformer  takes an input sequence and transforms it into a set of encoded vectors. These vectors encapsulate the important features and context of the input data like frequencies for the Fourier transform.

The decoder performs the inverse operation of the encoder; it takes the encoded vectors and transforms them back into a sequence that matches the format of the original input (in the case of NLP, into word embeddings), progressively predicting one more term of the sequence at each step. Following the previous analogy, the decoder can be thought of as a sort of inverse Fourier transform, where the frequencies are converted back to a time-domain signal.

\begin{figure}[h!!]
    \centering
    \includegraphics[width=0.5\textwidth]{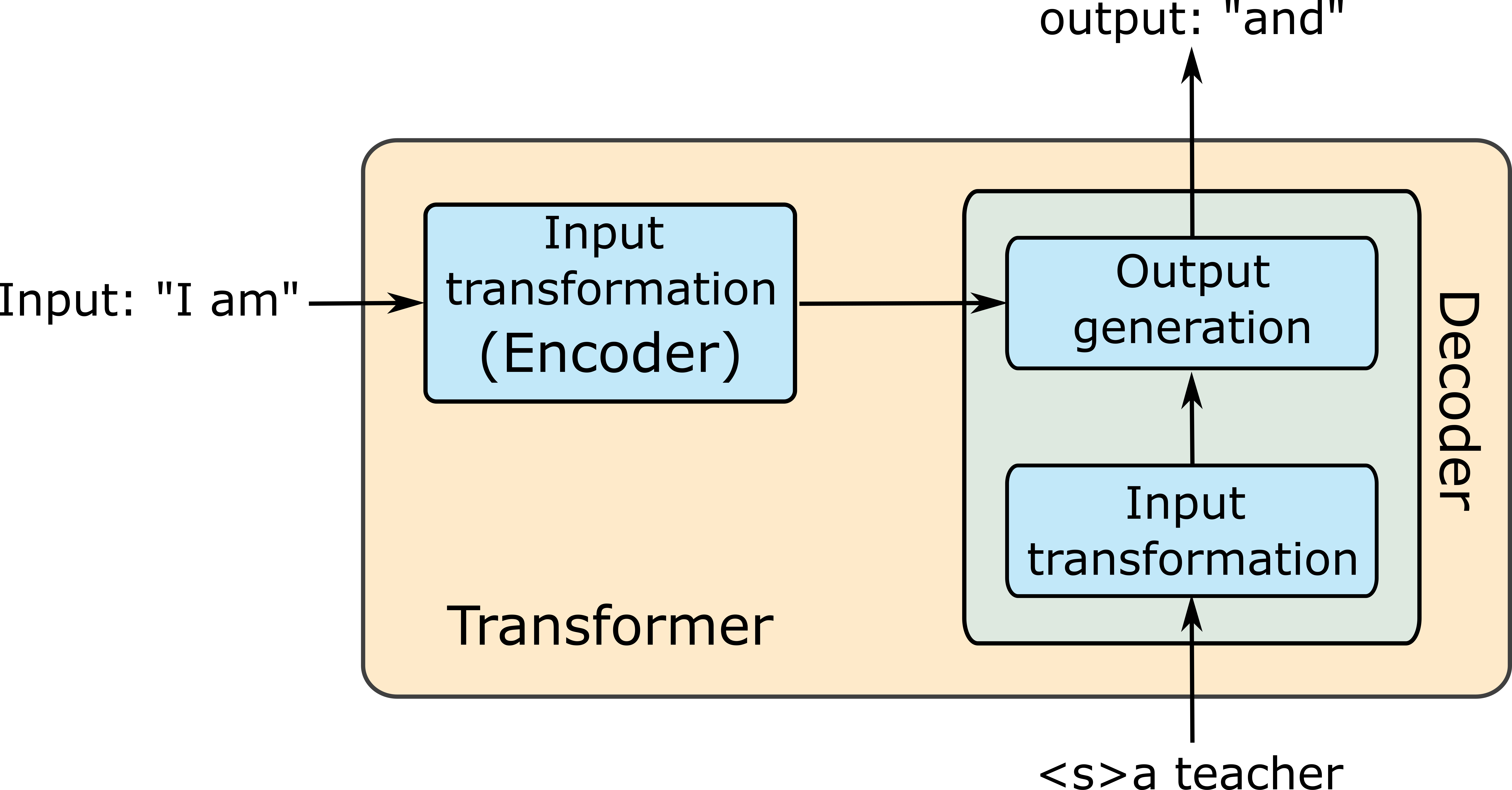}
    \caption{Encoder-Decoder architecture. Building on Figure~\ref{fig:encoder_decoder}, this figure shows how the decoder transforms its input sequence (\texttt{<s>} a teacher) into an internal representation before generating the next token ("and"). This process is similar to having an `encoder within the decoder,' as the input transformation in the decoder works in a way that is comparable to how the encoder processes the original input.}
    \label{fig:encoder_decoder2}
\end{figure}

One step further, we can divide the decoder into two main phases Figure~\ref{fig:encoder_decoder2}.
\break
The first phase is the input transformation phase, which processes the decoder input (e.g., $<$s$>$ a teacher in Figure ~\ref{fig:encoder_decoder2}). This phase converts the decoder input sequence in a way conceptually analogous to what the encoder does with the encoder input. We can think of it as an encoder within the decoder. Using the previous analogy, we can say that both user's and Transformer's inputs are transformed using a mathematical device conceptually similar to a Fourier transform. The transformations performed by the encoder and this part of the decoder are conceptually similar but not identical, as each one performs a different transformation that is determined during training.

After this transformation, the decoder combines the intermediate vectors derived from both the encoder and the first phase of the decoder. The second phase of the decoder is the output generation phase. In this phase, the decoder re-transforms the combined representations back into a sequence that matches the original input format, effectively predicting the next word or term in the sequence. Using our previous analogy, this is the phase that properly functions as a sort of inverse Fourier transform.

\subsection{Inside the ‘Input Transformation’ Box: Self-Attention Mechanism}
At the heart of the Transformer  architecture is the self-attention mechanism, which is central to the `Input Transformation' boxes in Figure~\ref{fig:encoder_decoder2}. This section focuses on the mathematical steps carried out by a self-attention layer during inference. For the moment, we will present these steps without providing any interpretation or context. While numerous introductory articles explain the reasoning behind these steps within the field of NLP (with many available on divulgative platforms like Towards Data Science, e.g., \cite{SelfAttention1_TowardsDataScience} \cite{SelfAttention2_TowardsDataScience}), our goal is to offer a practical introduction that extends beyond NLP. Initially, however, we will present only the bare mathematical steps.

\subsection*{Input Sequence}
We start with a sequence of input vectors $\bf{x}_i$ (for instance the embeddings of the user's input), where each vector $\bf{x}_i$ has a dimension of $d$. The sequence length is $N$, meaning there are $N$ such vectors in the sequence.

\subsection*{Linear Transformations}

Three linear transformations of the input sequence are carried by multiplying $\bf{x}_i$ with the matrices $\bm{W}_K$, $\bm{W}_Q$, and $\bm{W}_V$, each of size $d \times d'$. The transformed vectors $\bm{K}$, $\bm{Q}$, and $\bm{V}$ have dimension $d'$ and are called keys, queries and values. These names originate from terminology commonly used in the field of NLP, and while their specific origins are not relevant to our discussion, we adopt them here for consistency with existing literature. From our point of view, we can view these linearly transformed vectors as re-representations of the original sequence. These linear transformations, which may include rotation, scaling, shear, or projection of the original input data, preserve the fundamental properties of the original data while altering its dimensionality from $d$ to $d'$.

For each input vector $\mathbf{x}_{\mathit{i}}$, we calculate:
\begin{itemize}
  \item The key vector $\bm{K}_i = \bm{W}_K \mathbf{x}_{\mathit{i}}$
  \item The query vector $\bm{Q}_i = \bm{W}_Q \mathbf{x}_{\mathit{i}}$
  \item The value vector $\bm{V}_i = \bm{W}_V \mathbf{x}_{\mathit{i}}$
\end{itemize}
The numerical values of $\bm{W}_K$, $\bm{W}_Q$, and $\bm{W}_V$ are determined during training.

\subsection*{Compute Attention Scores}

For each pair of input vectors $(\bf{x}_i, \bf{x}_j)$ in the sequence, we define the attention score $s_{ij}$ as the dot product of the query vector $\bm{Q}_i$ of $\bf{x}_i$, and the key vector $\bm{K}_j$ of $\bf{x}_j$, rescaled by the square root of $d'$:
\begin{equation}
  s_{ij} = \frac{\bm{Q}_i \bm{K}_j}{\sqrt{d'}}
  \label{eq:sij}
\end{equation}

This results in a matrix of scores $\bm{s_{ij}} \in \mathbb{R}^{N \times N}$.

\subsection*{Apply Softmax to Obtain Attention Weights}

We apply the softmax function to each row of the score matrix to get the attention weights $\bm{S}_{ij}$:
\begin{equation}
  \bm{S}_{ij} = \frac{\exp(s_{ij})}{\sum_{k=1}^{N} \exp(s_{ik})}
  \label{eq:S_ij}
\end{equation}

The softmax ensures that the attention weights for each vector sum to 1, making them comparable to probabilities.

\subsection*{Compute Attention Output}

For each input vector $\bf{x}_i$, we compute the output vector $\bm{A}_i$ as the weighted sum of the value vectors $\bm{V}_j$, using the attention weights $\bm{S}_{ij}$:
\begin{equation}
  \bm{A}_i = \sum_{j=0}^{N-1} \bm{S}_{ij} \bm{V}_j
  \label{eq:Ai}
\end{equation}
The final output of the self-attention mechanism is a new sequence of vectors $\bm{A}_i$ of size $N \times d'$. Each output vector is influenced by the entire input sequence, weighted by their respective attention scores.

\subsection{The engineering perspective: Self-Attention Mechanism}

At first glance, the previous steps may seem quite arbitrary. To gain some intuition, we can relate these to similar concepts in physics and mathematics that are commonly familiar to engineers. We can think of Equation (\ref{eq:Ai}) as having the same form as a discrete integral transform where the sequence $\bm{V}_j$ is transformed into the sequence $\bm{A}_i$ by the kernel $\bm{S}_{ij}$. 
For comparison, a common integral transform, the discrete Fourier transform (DFT), can be written as:

\begin{equation}
X_k = \sum_{n=0}^{N-1} x_n e^{-i 2 \pi k n / N}
\end{equation}

In this expression, the time-sequence $x_n$ is transformed into the frequency-sequence $X_k$ by the kernel $e^{-i 2 \pi k n / N}$.
Equation  (\ref{eq:Ai}) is analogous to a Fourier series, but with a different kernel that is calculated from the data during training. Moreover, the ${s}_{ij}$ attention scores are the elements of the (centered) covariance matrix between $\bm{Q}_i$ and $\bm{K}_j$, just rescaled by $\sqrt{d'}$. But because $\bm{Q}_i$ and $\bm{K}_j$ are both linear transformations of the original sequence $\mathbf{x}_{\mathit{i}}$, the kernel $\bm{S}_{ij}$ provides the same information of an auto-covariance matrix.
Incidentally, an integral transform where the kernel is the autocorrelation matrix of the input sequence $\mathbf{x}_{\mathit{i}}$ is called a Karhunen-Loève transform \cite{Fontanella2012}, which is closely related to the principal component analysis (PCA) technique widely used in data analysis. Unlike a Fourier transform, where the kernel is predetermined, the kernel in the Karhunen-Loève transform is data-dependent, just as in Transformers.

\subsection{Inside the ‘Output Transformation’ Box: Cross-Attention Mechanism}
\begin{figure}[h!!]
    \centering
    \includegraphics[width=0.5\textwidth]{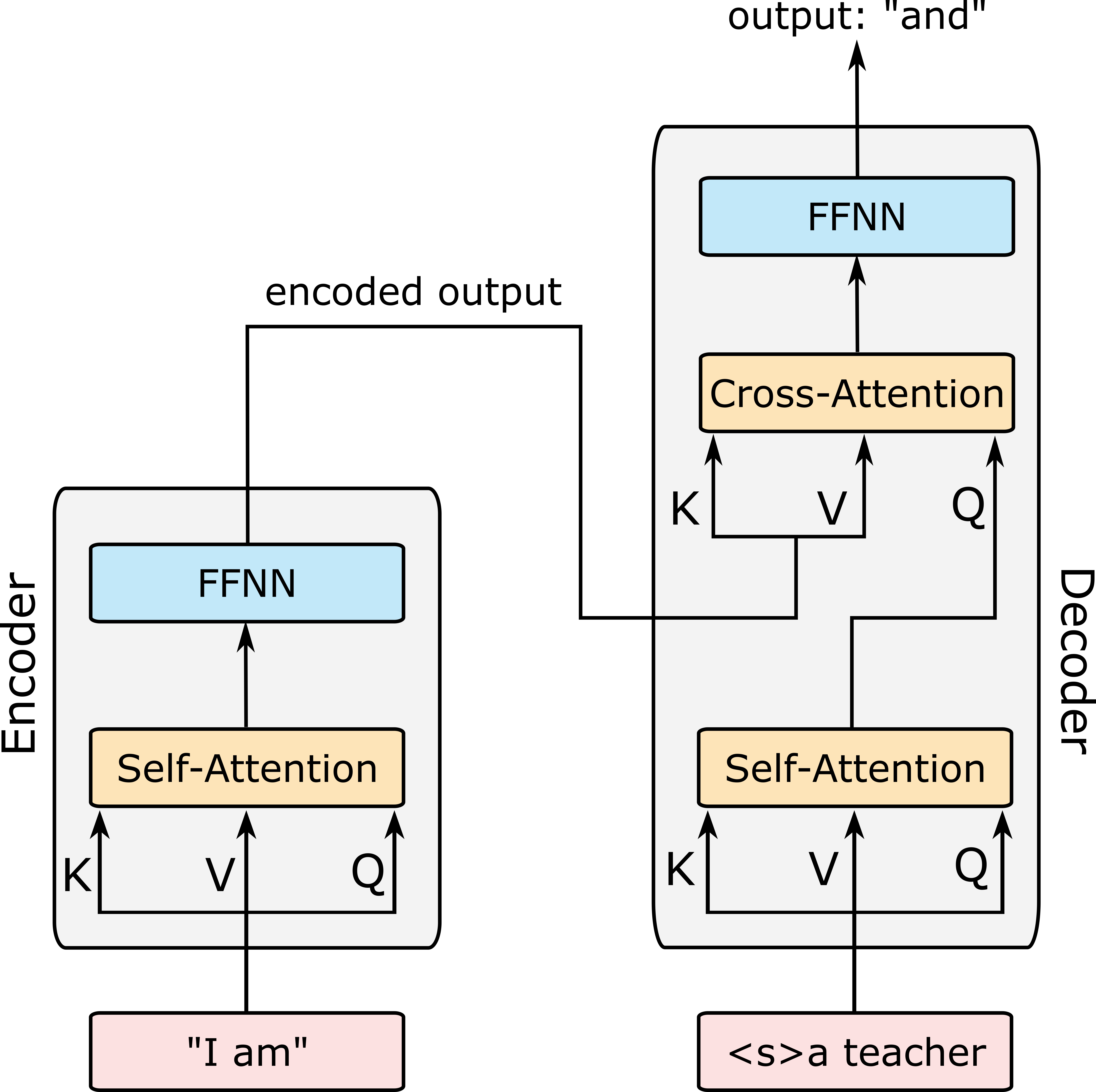}
    \caption{Self-attention and cross-attention. This figure expands on Figure~\ref{fig:encoder_decoder2} by `opening up' the Input Transformation and Output Generation boxes to show the self-attention mechanism inside the encoder and decoder, and the cross-attention mechanism inside the decoder. Self-attention helps capture dependencies within the input, while cross-attention allows the decoder to focus on relevant parts of the encoder's output.}
    \label{fig:cross_attention}
\end{figure}

Let us delve one step deeper into the Transformer  architecture (Figure~\ref{fig:cross_attention}). From the previous section, we know that both the encoder and decoder inputs are transformed into keys, queries, and values through linear transformations.
\break

These are then processed by the self-attention layers to produce attention outputs. In the cross-attention mechanism within the decoder, the keys and values are derived from the encoder's output, while the queries come from the decoder's self-attention outputs.
This means that the decoder attends to the encoder's outputs, allowing it to incorporate information from the entire input sequence while generating each output token.
Mathematically, cross-attention works in the same way of self-attention (\ref{eq:sij} to \ref{eq:Ai}) with the difference that now $\bm{Q}$ comes from the decoder while $\bm{K}$ and $\bm{V}$ from the encoder.

Figure~\ref{fig:cross_attention} also shows the presence of Feed Forward Neural Networks (FFNNs) after the self-attention layer. These FFNNs are crucial for the functioning of the Transformer  as they introduce non-linearity and increase the model's capacity by adding more trainable parameters.

\subsection{The Engineering Perspective: Cross-Attention Mechanism}

In line with our previous analogy to the Fourier transform, we can see the cross-attention mechanism as a specialized transformation that utilizes the interactions between two different sets of representations, just as self-attention does with one set. In fact, self-attention captures relationships within the same set of inputs, while cross-attention combines information from the encoder's output with the decoder's queries.

While self-attention can be viewed as a form of auto-correlation, identifying relationships within the same input sequence, cross-attention operates more like a cross-correlation. In this context, the kernel used in cross-attention reflects the dependencies between the output of the encoder (serving as keys and values) and the intermediate representations in the decoder (serving as queries).

To draw an engineering analogy, think of the encoder's output as a complex signal that has been transformed into a frequency domain representation. The decoder's queries then act as a set of filters that selectively extract and recombine specific frequency components from this signal to reconstruct the desired output sequence. This process is similar to how engineers might use cross-correlation techniques in signal processing to compare and align different signals for analysis or reconstruction.

Mathematically, the continuous cross-correlation of two continuous functions \( f \) and \( g \) is given by:
\begin{equation}
(f \star g)(t) = \int_{-\infty}^{\infty} f(\tau) g(\tau + t) \, d\tau
\end{equation}

In practice, for discrete signals, the cross-correlation is often computed over a finite range:
\begin{equation}
(f \star g)[n] = \sum_{m=0}^{N-1} f[m] \cdot g[m+n]
\end{equation}
where \( N \) is the length of the signals.

Cross-correlation is used to measure the similarity between two signals as a function of the time-lag applied to one of them. For instance, in radar signal processing, cross-correlation helps detect the presence of a target by comparing the received signal with a known transmitted signal. The peak in the cross-correlation function indicates the time delay at which the signals best align, revealing the target's location.

Applying this analogy to Transformers,  consider the encoder’s output as the reference signal and the decoder’s queries as the signal being shifted. Cross-attention computes how much each part of the encoder’s output (keys and values) should influence the current part of the decoder's output (queries). The resulting attention weights indicate the degree of relevance or similarity, guiding the integration of information from the encoder to the decoder.

\subsection{Multi-Headed Attention}
\label{sec:multiheaded_attention}

In the previous sections, we discussed the self-attention mechanism and its analogy to integral transforms, such as the Fourier transform. We saw how a single self-attention layer can transform an input sequence into a set of weighted representations. Now, we will introduce the concept of multi-headed attention, which can be understood as performing multiple, parallel transformations on the input data, each potentially capturing different aspects of the underlying structure.

Multi-headed attention can be thought of as performing several distinct Fourier-like transforms on the input data, each one with a different kernel learned from the data during the training process. Each head in multi-headed attention applies its own set of linear transformations to the input sequence, generating unique sets of keys, queries, and values. This results in different attention outputs for each head, enabling the model to capture a richer set of relationships in the data. 

For example, one head might capture short-term dependencies in the data, analogous to detecting high-frequency components in a Fourier transform, while another might focus on long-term dependencies, similar to identifying low-frequency components. However, this comparison to the Fourier transform is just a simplification to help engineers understand the concept. In reality, each head is likely to capture more complex and nuanced patterns that are not easily interpreted through this simple analogy.

The outputs from the different heads are combined to produce the final representation. This is typically done in the following sequential steps:
\begin{enumerate}
\item \textbf{Concatenation}: The output vectors from all heads are concatenated along the feature dimension. If there are $h$ heads and each head produces an output of dimension $d'$, the concatenated output will have a dimension of $h \times d'$.
\item \textbf{Linear Transformation}: The concatenated output is then linearly transformed using a weight matrix $\bm{W}_O$ to produce the final output of the multi-headed attention layer. This step ensures that the combined output has the desired dimension for subsequent layers.
\end{enumerate}

Mathematically, if $\bm{A}_i^{(j)}$ is the output of the $j$-th head for the $i$-th input, the combined output $\bm{A}_i$ is:
\begin{equation}
    \bm{A}_i = \bm{W}_O \left[ \bm{A}_i^{(1)} \| \bm{A}_i^{(2)} \| \cdots \| \bm{A}_i^{(h)} \right]
\end{equation}
where $\|$ denotes concatenation.

Thus, Figure~\ref{fig:cross_attention} can be further developed by substituting the self- and cross-attention blocks with multi-headed (self- and cross-attention) blocks.

\subsection{Positional embeddings}
\begin{figure}[h!!]
    \centering
    \includegraphics[width=0.25\textwidth]{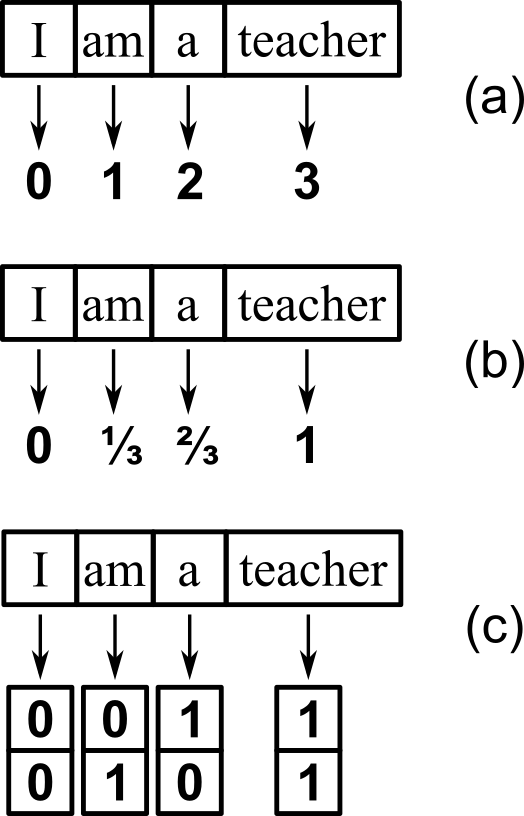}
    \caption{Ineffective methods for adding positional information to token embeddings. (a) Index-based encoding assigns a simple index to each token (e.g., 0, 1, 2, 3), which can lead to large gradients at higher positions. (b) Normalized index encoding scales positional values between 0 and 1 by dividing each index by the sequence length, but it may cause ambiguity across sequences of different lengths. (c) Binary encoding represents positions using fixed-length binary vectors, but lacks smoothness in positional transitions.}
    \label{fig:positions}
\end{figure}

Parallel processing is a key strength of Transformers,  but it also means they don't inherently understand the concept of order. For example, the attention mechanism can't differentiate between the sentence "I am a teacher" and its shuffled version "teacher a am I." This limitation reinforces our earlier Fourier analysis analogy. In fact, Fourier analysis breaks down a signal (like a piece of music) into its constituent frequencies (notes) without indicating when those notes are played. This tells you what notes are in the piece but not their order or timing. In signal analysis, this limitation is addressed by wavelets, which effectively add the dimension of time to the frequency analysis, making them more suitable for analyzing signals where timing and order are crucial, like music. Similarly, in Transformers,  this is achieved by adding positional embeddings to the input tokens, allowing the model to capture the order of elements in a sequence. There are many different approaches that have been used to explain positional embeddings in a simple way. Here, we follow the approach of \cite{Positional_Encodings_TowardsDataScience2} and \cite{Positional_Encodings_TowardsDataScience}.\\ 
A naive approach might be to simply add a positional encoding where each entry is its index number Figure~\ref{fig:positions}a. However, this solution is not ideal because it assigns higher numerical values to tokens at later positions, potentially leading to issues like exploding gradients. Therefore, positional embeddings need to be normalized. Dividing by the largest integer would rescale values in [0,1] Figure~\ref{fig:positions}b, but creates a problem with arbitrary sequence lengths: a value of 0.5 means different things in sequences of varying lengths. To address this, one could convert positions to binary, associating each position with a vector of 0s and 1s Figure~\ref{fig:positions}c. This normalizes the values, but these binary vectors are not "smooth" since they only take values of 0 or 1. Smoothness is crucial for gradient-based optimization because it ensures that small changes in input lead to small changes in output, allowing effective learning during training. To resolve this, we need a continuous version of these binary vectors.
\\
\break
As an example, let us consider a sequence of length $l_{seq}$ = 16 tokens. Therefore, each position, denoted as $pos$ corresponds to a binary vector of dimension 4:
\begin{center}
\begin{tcolorbox}[colback=gray!20, colframe=gray!40, boxrule=0.5mm, width=0.25\textwidth]
0  :  0000 \\
1  :  0001 \\
2  :  0010 \\
3  :  0011 \\
4  :  0100 \\
5  :  0101 \\
6  :  0110 \\
7  :  0111 \\
8  :  1000 \\
9 :  1001 \\
10 :  1010 \\
11 :  1011 \\
12 :  1100 \\
13 :  1101 \\
14 :  1110 \\
15 :  1111
\end{tcolorbox}
\end{center}

\begin{figure}[h!!]
    \centering
    \includegraphics[width=0.5\textwidth]{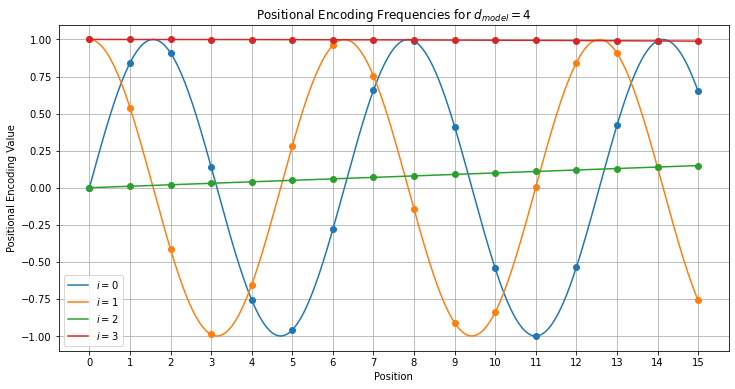}
    \caption{Frequencies of positional encodings generated using sine and cosine functions. Each dimension of the positional encoding captures a different frequency, with smaller dimensions representing higher frequencies and larger dimensions corresponding to lower frequencies.}
    \label{fig:PE_Frequencies}
\end{figure}

We denote the dimension of these binary vectors as $d_{model}$, and the index within this vector as $i$.\\ 
To develop an intuition for positional encoding, let us observe how the pattern of bits changes across columns: the least significant bit (rightmost) toggles with every increment, corresponding to a frequency of \( \frac{1}{2} \). The second bit toggles every two increments, with a frequency of \( \frac{1}{4} \). The third bit toggles every four increments, reflecting a frequency of \( \frac{1}{8} \), and so forth. This pattern of different frequencies is the key to understanding positional encoding. Instead of using discrete bits, we can define positional embedding $PE$ using smoother functions, like sine and cosine functions.
\begin{equation}
PE_{(pos,i)} = 
\begin{cases} 
\sin\left(\frac{pos}{10000^{i/d_{model}}}\right) & \text{if } i \text{ is even}\\
\cos\left(\frac{pos}{10000^{(i-1)/d_{model}}}\right) & \text{if } i \text{ is odd}
\end{cases}
\end{equation}
The term \( 10000^{i/d_{model}} \) in the formula serves to decrease the frequency as \( i \) becomes larger. Smaller \( i \) values represent the least significant bits of the binary number that change with higher frequency, while larger \( i \) values correspond to bits that change with lower frequency. Now the first 16 binary number become
\\
\begin{center}
\begin{tcolorbox}[colback=gray!20, colframe=gray!40, boxrule=0.5mm, width=0.7\textwidth]
\begin{tabular}{r r r r r r}
0  & : & \( 1.000 \) & \( 0.0000 \) & \( 1.0000 \) & \( 0.0000 \) \\
1  & : & \( 0.9999 \) & \( 0.0099 \) & \( 0.5403 \) & \( 0.8415 \) \\
2  & : & \( 0.9998 \) & \( 0.0199 \) & \(-0.4161 \) & \( 0.9093 \) \\
3  & : & \( 0.9995 \) & \( 0.0299 \) & \(-0.9900 \) & \( 0.1411 \) \\
4  & : & \( 0.9992 \) & \( 0.0399 \) & \(-0.6536 \) & \(-0.7568 \) \\
5  & : & \( 0.9987 \) & \( 0.0499 \) & \( 0.2837 \) & \(-0.9589 \) \\
6  & : & \( 0.9982 \) & \( 0.0599 \) & \( 0.9602 \) & \(-0.2794 \) \\
7  & : & \( 0.9975 \) & \( 0.0699 \) & \( 0.7539 \) & \( 0.6570 \) \\
8  & : & \( 0.9968 \) & \( 0.0799 \) & \(-0.1455 \) & \( 0.9894 \) \\
9 & : & \( 0.9959 \) & \( 0.0898 \) & \(-0.9111 \) & \( 0.4121 \) \\
10 & : & \( 0.9950 \) & \( 0.0998 \) & \(-0.8391 \) & \(-0.5440 \) \\
11 & : & \( 0.9939 \) & \( 0.1097 \) & \( 0.0044 \) & \(-1.0000 \) \\
12 & : & \( 0.9928 \) & \( 0.1197 \) & \( 0.8439 \) & \(-0.5366 \) \\
13 & : & \( 0.9915 \) & \( 0.1296 \) & \( 0.9074 \) & \( 0.4202 \) \\
14 & : & \( 0.9902 \) & \( 0.1395 \) & \( 0.1367 \) & \( 0.9906 \) \\
15 & : & \( 0.9887 \) & \( 0.1494 \) & \(-0.7597 \) & \( 0.6503 \) \\
\end{tabular}
\end{tcolorbox}
\end{center}

which correspond to the frequencies in Figure~\ref{fig:PE_Frequencies}. There are additional properties of Positional Encodings defined in this way, as well as alternative definitions, which are not discussed here. Interested readers can refer to divulgative articles such as 
\cite{Positional_Encodings_TowardsDataScience}.
\\
\break

\subsection{The final touches (inference)}
\begin{figure}[h!!]
    \centering
    \includegraphics[width=0.7\textwidth]{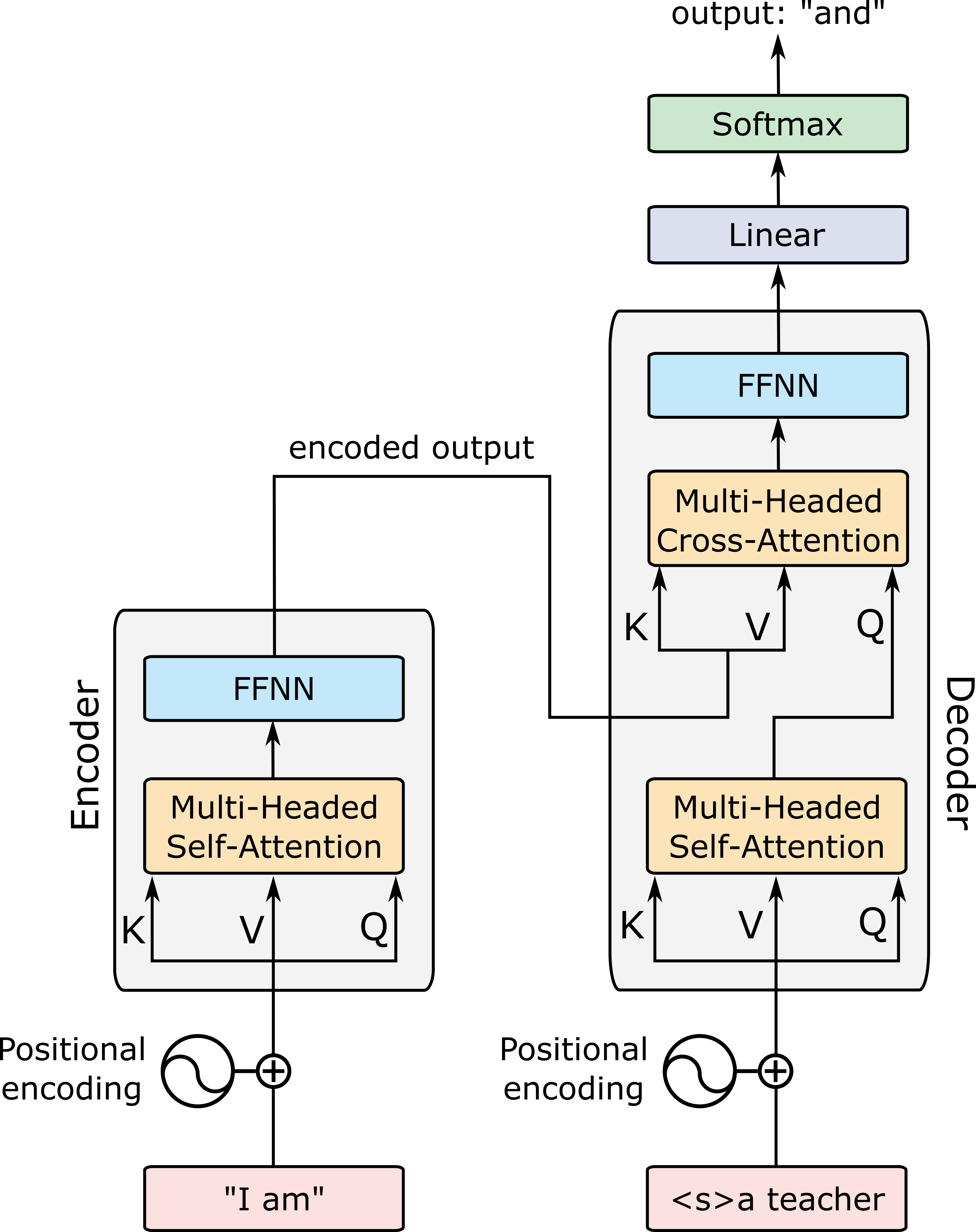}
    \caption{Complete inference process in Transformers. Building on Figure~\ref{fig:cross_attention}, positional encodings are added to the input tokens, multi-headed self-attention and cross-attention mechanisms are introduced, and feed-forward neural networks (FFNNs) are applied. The final output is generated through a linear layer followed by a Softmax function.}
    \label{fig:final_inference}
\end{figure}

Several additional layers complete the Transformer  architecture during inference (Figure~\ref{fig:final_inference}). First, a Linear layer projects the output embeddings from the decoder into a higher-dimensional space corresponding to the vocabulary size. Each element in this vector represents an unnormalized score (logit) for a specific word being the next token in the sequence. Finally, a Softmax layer converts these logits into probabilities, indicating the likelihood of each word being the correct next token.\\

\break

\subsection{Training}
So far, we have discussed how the Transformer model operates during inference, where it predicts the next word in a sequence based on a given input from the user. However, for the model to generate coherent and contextually appropriate text during inference, it must first undergo a training process. A Transformer model contains a large number of trainable parameters, including those within the FFNNs and the various weight matrices $\bm{W}_K$, $\bm{W}_Q$, and $\bm{W}_V$ that are used in the self-attention mechanism. These parameters must be trained (i.e., optimized) so the model can effectively predict the next word in a sequence given all the previous words. We assume the reader is familiar with fundamentals of training neural networks such as back-propagation and Stochastic Gradient Descent.\\
Let’s revisit the sequence from our earlier example: the user's input was "I am," and the model recursively generated the output <s> a teacher and you are a student <eos>."
In practice, the training process closely mirrors the inference process. For example:

\begin{itemize}
    \item Given the user input "I am" and the Transformer input "<s>", the target output for training is "a".
    \item Then, given the user input "I am" and the Transformer input "<s> a", the target output for training is "teacher".
    \item This process continues similarly for each subsequent word in the sequence.
\end{itemize}

\begin{figure}[h!!]
    \centering
    \includegraphics[width=0.5\textwidth]{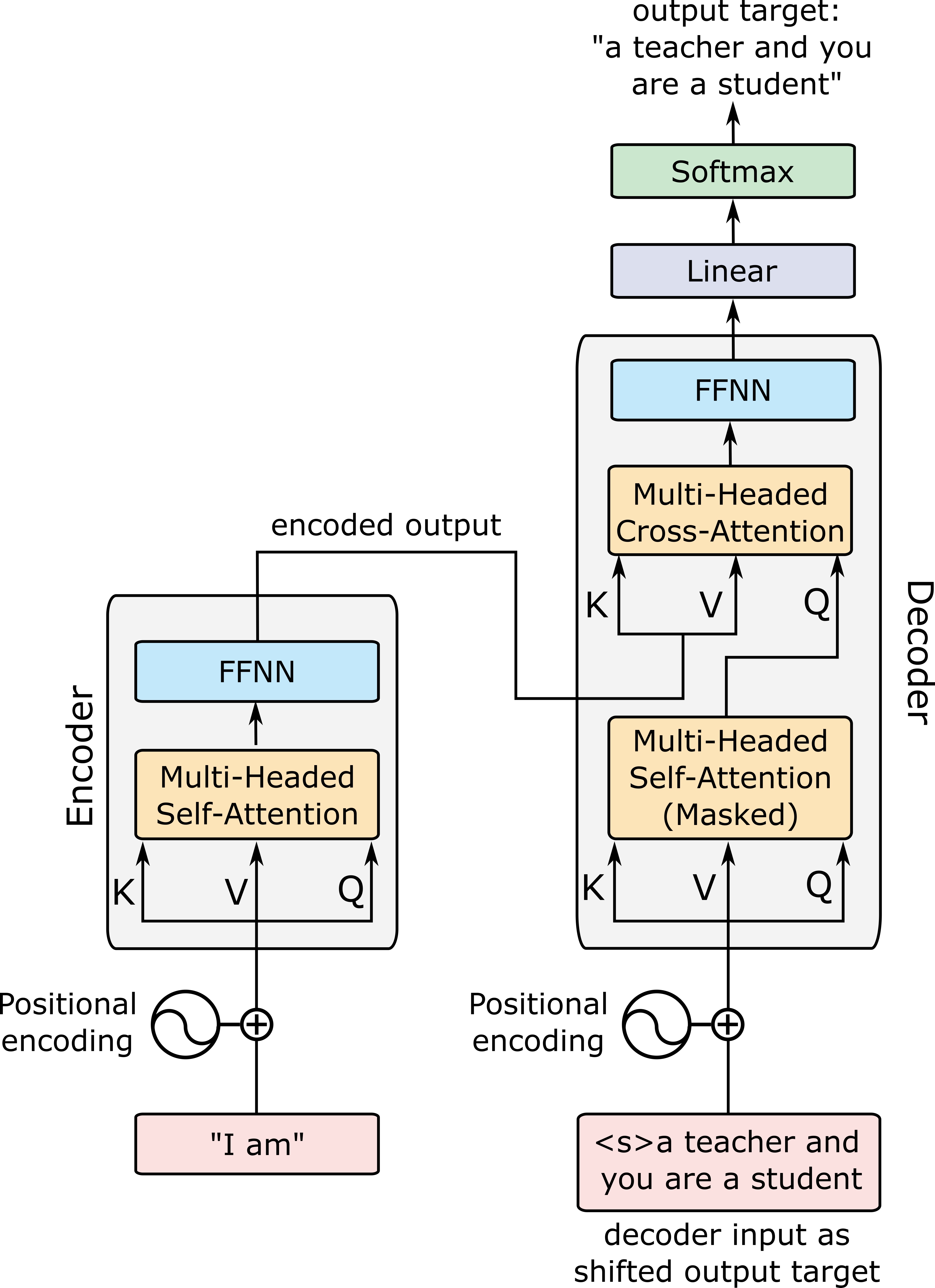}
    \caption{Training the Transformer. This figure illustrates the training process, where the entire input sequence (`<s> a teacher and you are a student') is fed into the decoder simultaneously. The target sequence (`a teacher and you are a student <eos>') is used to guide learning.}
    \label{fig:almost_complete}
\end{figure}

However, there is a key difference between the conceptual description above and the actual training process of Transformers, which is designed to take advantage of parallelization. Instead of processing the sequence word by word (which would be sequential and slow), the entire sequence "<s> a teacher and you are a student" is fed into the decoder all at once during training. Simultaneously, the entire sequence "a teacher and you are a student <eos>" is used as the target output (Figure~\ref{fig:almost_complete}). However, despite this, the model still learns to associate the first element of the decoder input, "<s>", with the first element of the target output, "a", the second element "a" with the second target "teacher", and so on.\\

\break

\begin{figure}[h!!]
    \centering
    \includegraphics[width=0.7\textwidth]{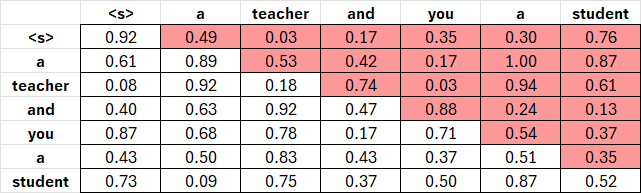}
    \caption{Mask applied to the decoder during training. The red cells represent the positions in the attention matrix that are masked, ensuring that the decoder does not see future tokens during training.}
    \label{fig:mask}
\end{figure}

There is one challenge, however. During the training process, the multi-headed attention mechanism in the decoder would, by default, calculate attention scores across all combinations of the decoder input sequence, "<s> a teacher and you are a student" (Figure~\ref{fig:mask}).

The problem arises when the model tries to learn from the decoder input rather than the target output. For example, when associating "<s>" with "a", the decoder's self-attention should not allow "a" (or any successive word) to influence this prediction since "a" is the correct answer the model is supposed to learn from the target. Similarly, when the model is learning to associate "a" with "teacher", it should not be influenced by "teacher" or any subsequent words, because it shouldn't know that "teacher" is the next word.

To prevent the model from "cheating" by using future words from its attention scores, a mask is applied to the self-attention mechanism of the decoder during training. The attention scores in red in (Figure~\ref{fig:mask}) are effectively removed by applying this mask.

The mask is a matrix of the same size as the attention scores, filled with values of 0's and negative infinities. When this mask is added to the scaled attention scores, it replaces the upper triangular portion (the future positions) with negative infinities. This modification ensures that, when the softmax function is applied to these scores, the negative infinities are zeroed out, meaning the model does not attend to future tokens and thus cannot use them to make its predictions.

\break

\subsection{The final touches (training)}
\begin{figure}[h!!]
    \centering
    \includegraphics[width=0.5\textwidth]{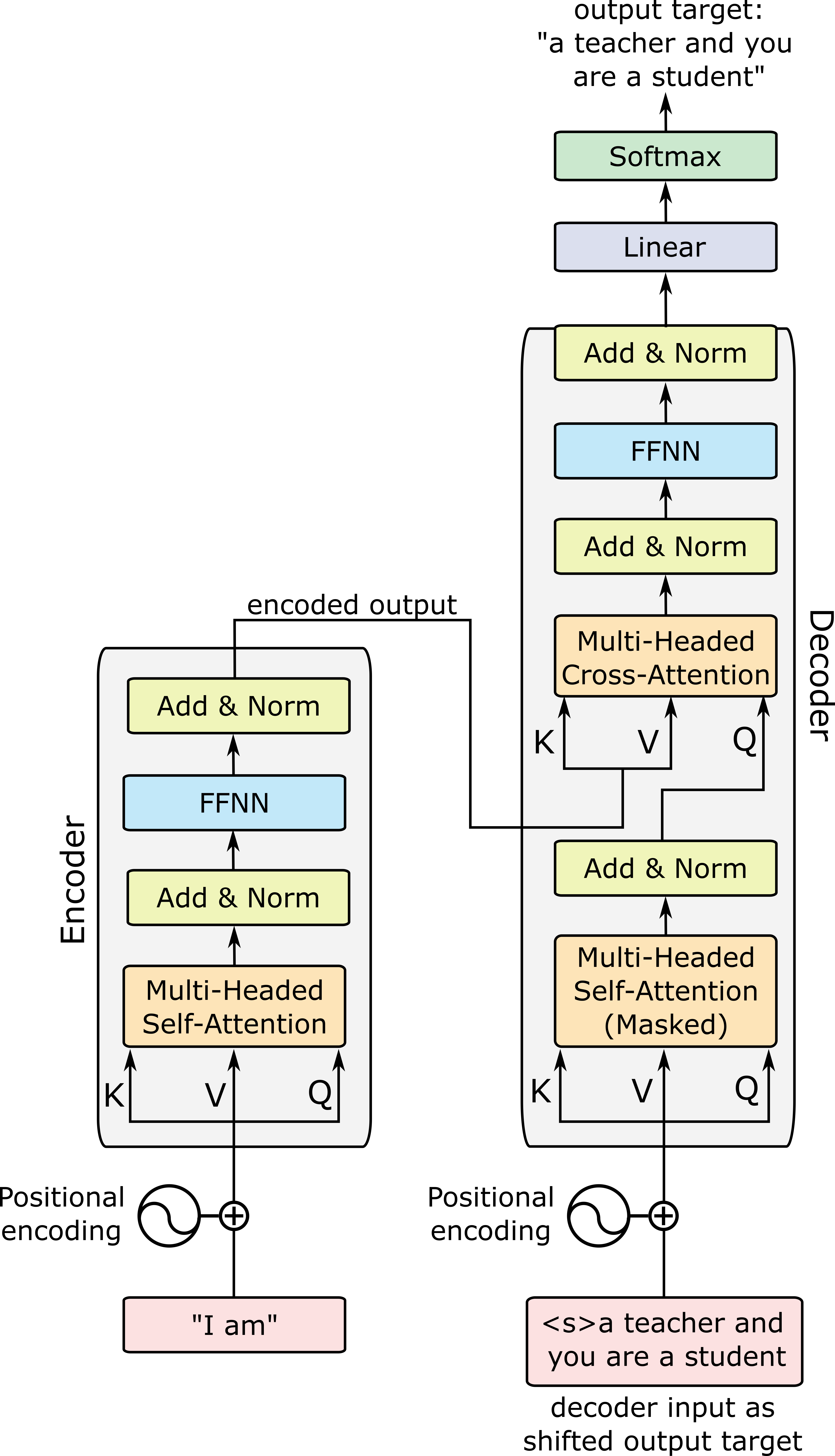}
    \caption{After adding the final Add \& Norm layers, which combine residual connections and normalization to stabilize training, we achieve the full Transformer architecture in its complete form.}
    \label{fig:complete}
\end{figure}

Finally, Figure~\ref{fig:complete} includes several "Add \& Norm" layers, which are used to improve training.  The "Add" operation represents a residual connection, where the output of a sub-layer is added to its original input. This helps mitigate the vanishing gradient problem during backpropagation. The "Norm" operation refers to Layer Normalization, which normalizes the output by rescaling it based on the mean and variance of the layer’s activations. This normalization stabilizes the training process by ensuring consistent activation scales.

After this in-depth introduction to Transformers, we can conceptualize the entire process as an advanced exercise in multidimensional fitting, facilitated by two key types of transformations. The first type is explicit function mapping, executed by FFNNs, where each specific set of input values is directly mapped to a corresponding output set. The second type involves global mapping through the attention mechanism, which functions similarly to a (discrete) integral transform. This process integrates a function over the entire sequence with a learned kernel, producing new variables or functions that depend on the sequence as a whole. Unlike fixed transformations, such as the Fourier transform, these fitting functions are dynamically learned from data during training, allowing the model to adapt and refine its understanding.

With this understanding, we can now move to the second part of the paper, where we apply these insights to use Transformers for learning Partial Differential Equations (PDEs). This next section will demonstrate how the principles of multidimensional fitting discussed here can be extended to tackle engineering problems beyond NLP.

\break

\section{Methods}
\label{sec:Methods}

As we have seen, even when applied in their native NLP domain, Transformers translate sequences of letters and words into numerical representations (embeddings). We will leverage this inherent feature of Transformers to train a model to predict the evolution of temperature in a heat conduction problem.

\subsection{The physical problem: basic configuration}
\label{sec:sub:MethodsBasic}
\begin{figure}[h!!]
    \centering
    \includegraphics[width=0.5\textwidth]{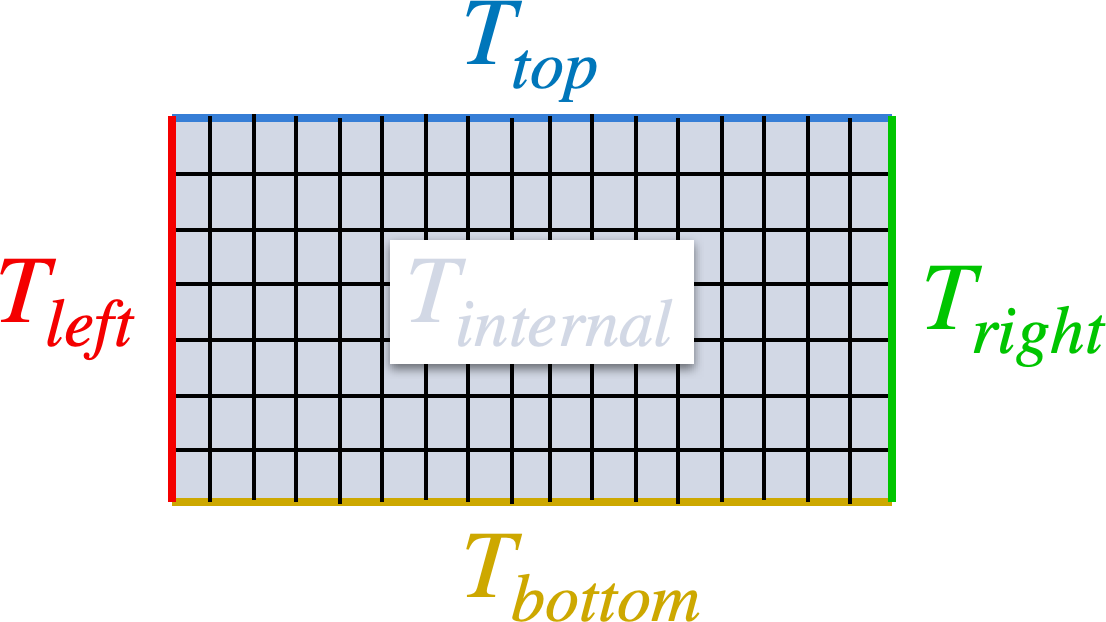}

    \caption{The domain for the basic problem setup with four Dirichlet boundary conditions.}
    \label{fig:domain}
\end{figure}

First, let's make clear what exactly we're training the Transformer  to predict. We consider a 2D plate with four Dirichlet BCs and an initial temperature distribution Figure~\ref{fig:domain}. We use finite differences to solve the problem multiple times, each with randomly selected BCs, initial state and thermal diffusivity. The resulting finite-difference solutions provide separate training, validation and test sets for the Transformer model. For each case that is solved with finite differences, the normalized temperature on the left, top and right sides of the plate is randomly chosen in the normalized range 0 to 1.0. At the bottom side the normalized temperature is randomly chosen in the range 0 to 0.10 to induce further asymmetry to the problem. Furthermore, the initial internal temperature for each run is also initialized to a value that is randomly chosen in the range 0 to 1, independently from the Dirichlet temperatures. Finally, the heat diffusivity is also randomly chosen within a specified range. Because the heat diffusivity changes, the time step imposed by the stability criteria for the finite differences also potentially differ from one run to the next.

\break

\subsection{The physical problem: challenge configuration 1}
\label{sec:sub:MethodsChallenge1}
\begin{figure}[h!!]
    \centering
    \begin{subfigure}[b]{\textwidth}
        \centering
        \includegraphics[width=0.9\textwidth]{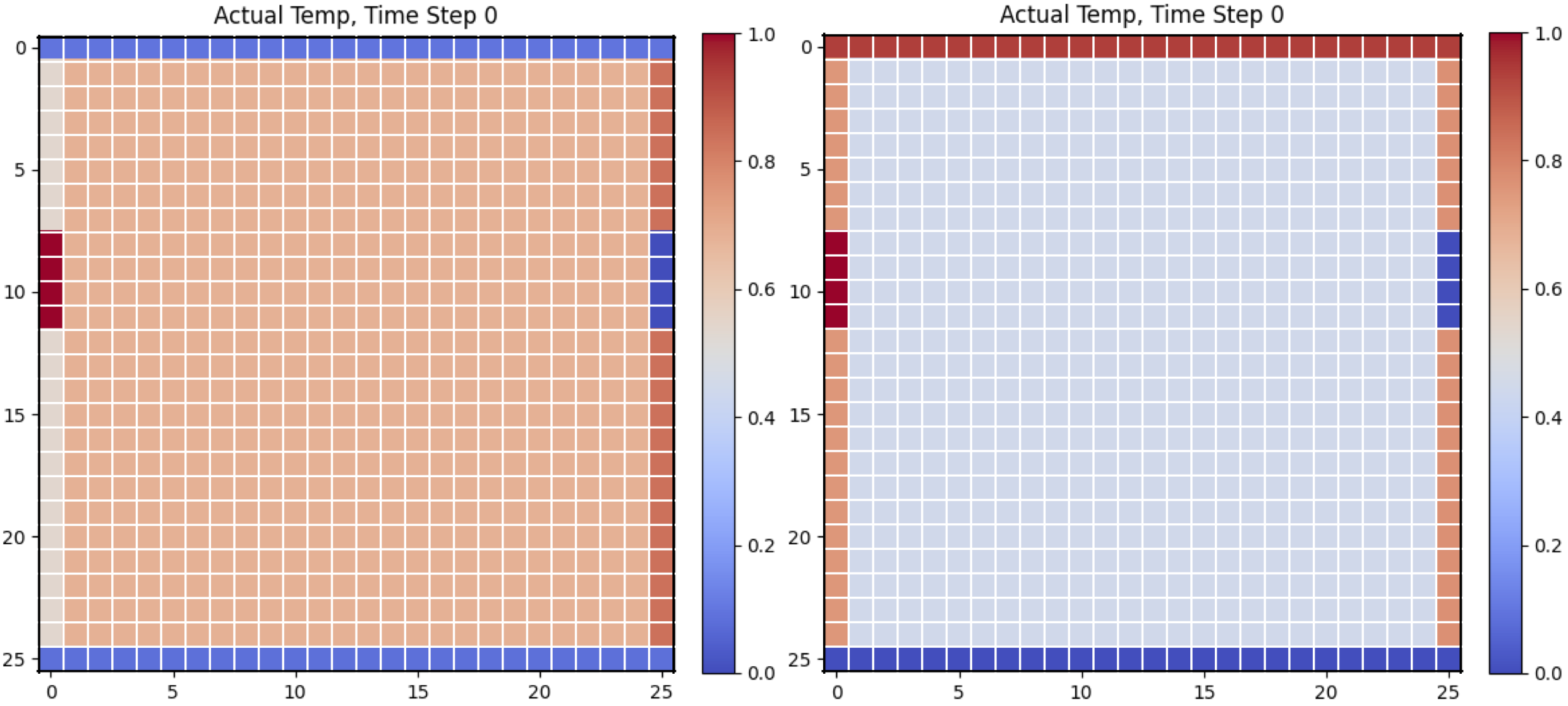}
        \caption{}
        \label{fig:domain_challenge1_a}
    \end{subfigure}
    \hfill
    \begin{subfigure}[b]{\textwidth}
        \centering
        \includegraphics[width=0.9\textwidth]{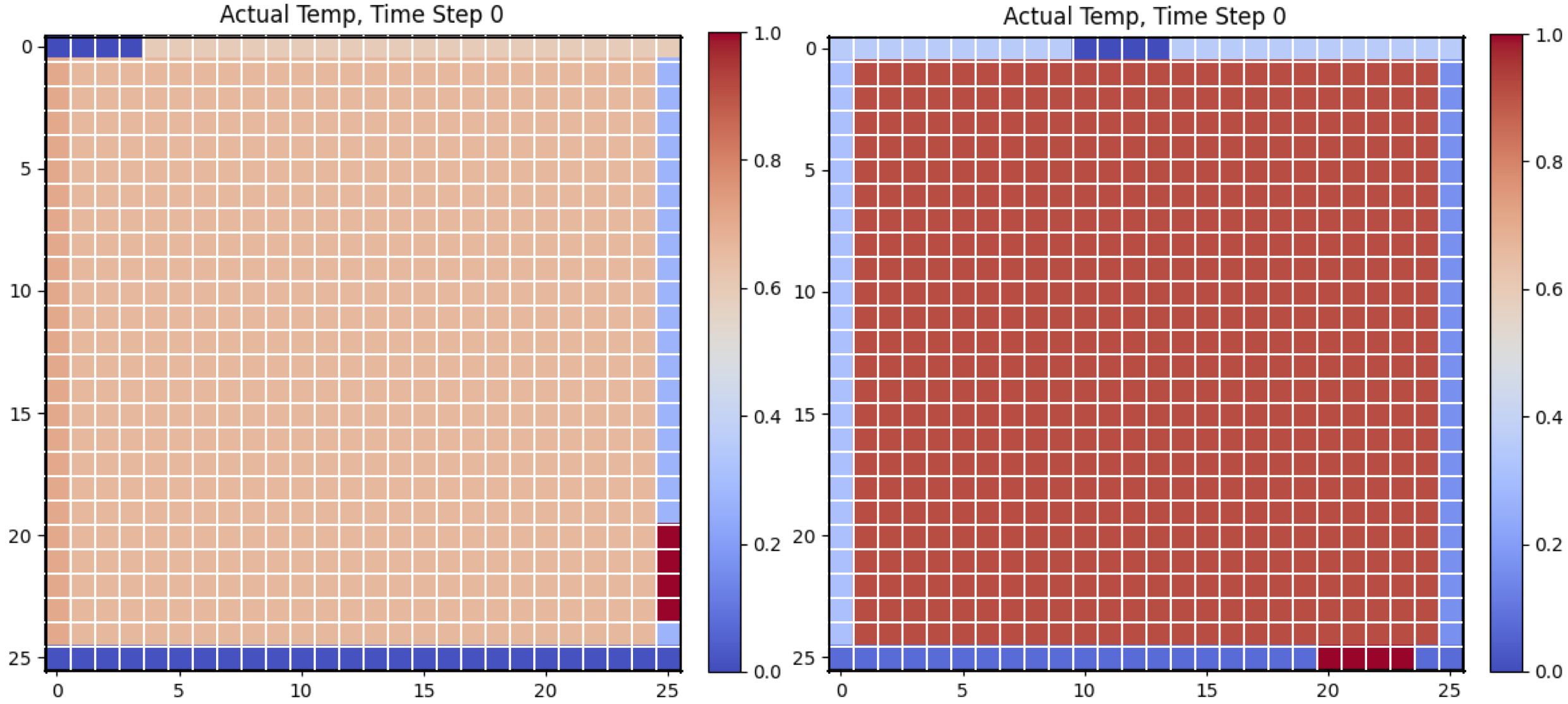}
        \caption{}
        \label{fig:domain_challenge1_b}
    \end{subfigure}
    \caption{(a) Two examples of the domain configuration for Challenge 1 with four Dirichlet boundary conditions and two fixed boundary segments; (b) Two examples of the domain configuration for  Challenge 2 with four Dirichlet boundary conditions and two randomly placed boundary segments.}
    \label{fig:domain_challenge1and2}
\end{figure}

We also consider a configuration that is designed to challenge the model by increasing the range of variability that the model must learn to adapt to. We consider two segments of predefined length, one with normalized temperature T=1 and the second with T=0, that are placed at specified locations on the left and right boundaries respectively as shown in Figure~\ref{fig:domain_challenge1and2}a.  The introduced complication increases the  likelihood that during inference the model will encounter conditions that deviate more substantially from those it had encountered during training. The reason is that these boundary features cause regions of localized inhomogeneity in the temperature field during the evolution towards steady state. Furthermore, both the extent of the associated inhomogeneities and how quickly they spread away from the boundary segments depends both on the Dirichlet boundary conditions, the initial internal temperature distribution and the thermal diffusivity, which in the test set will be different than those encountered by the model during training.  We also wanted to test the effect of the additional boundary features on the weights of the final projection layer to understand how the model learns to emphasize features.

\subsection{The physical problem: challenge configuration 2}
\label{sec:sub:MethodsChallenge2}
Finally, we consider an even more advanced challenge designed to test the model's ability to generalize. We again consider two segments of predefined length, one with normalized temperature T=1 and the second with T=0.  Now however for each case in the dataset, the two segments are randomly placed on any of the Dirichlet boundaries with the requirement that the segments must occupy different sides of the plate (see Figure~\ref{fig:domain_challenge1and2}b). The  introduced complication increases even further the likelihood that during inference the model will encounter conditions that deviate  substantially from those it had encountered during training. 

\break
\subsection{Finite Differences Discretization}
\label{sec:sub:MethodsDiscretization}
While the heat diffusivity varies across cases in the dataset, it remains constant during the evolution of the temperature field. Thus, we introduce a non-dimensional thermal diffusivity parameter, \( \beta \), which is the ratio of the physical diffusivity to a reference diffusivity. The non-dimensionalized two-dimensional heat conduction equation is then given by,

\begin{equation}
\frac{\partial \theta}{\partial \tau} = \beta \left( \frac{\partial^2 \theta}{\partial \xi^2} + \frac{\partial^2 \theta}{\partial \eta^2} \right)
\end{equation}

\noindent where \( \theta = \frac{T - T_{\text{ref}}}{T_0} \) is the non-dimensional temperature, \( \tau \) is the non-dimensional time, and \( \xi \), \( \eta \) are the non-dimensional spatial coordinates corresponding to \( x \) and \( y \).

The spatial domain, a rectangular plate, is discretized into a uniform grid with \( N_{\xi} \) and \( N_{\eta} \) nodes along the \( \xi \) and \( \eta \) directions, respectively. The grid spacing is \( \Delta \xi = \frac{1}{N_{\xi}-1} \) and \( \Delta \eta = \frac{1}{N_{\eta}-1} \).

The central finite difference method is used to approximate the spatial derivatives. The second-order derivatives are approximated as follows,

\begin{equation}
\frac{\partial^2 \theta}{\partial \xi^2} \approx \frac{\theta_{i+1,j} - 2\,\theta_{i,j} + \theta_{i-1,j}}{(\Delta \xi)^2}
\end{equation}

\begin{equation}
\frac{\partial^2 \theta}{\partial \eta^2} \approx \frac{\theta_{i,j+1} - 2\,\theta_{i,j} + \theta_{i,j-1}}{(\Delta \eta)^2}
\end{equation}

\noindent Combining these, the non-dimensional heat conduction equation in discrete form is,

\begin{equation}
\frac{d\theta_{i,j}}{d\tau} = \beta \left( \frac{\theta_{i+1,j} - 2\,\theta_{i,j} + \theta_{i-1,j}}{(\Delta \xi)^2} + \frac{\theta_{i,j+1} - 2\,\theta_{i,j} + \theta_{i,j-1}}{(\Delta \eta)^2} \right)
\end{equation}

\noindent For the time integration, the explicit Euler method is used:

\begin{equation}
\theta_{i,j}^{n+1} = \theta_{i,j}^n + \Delta \tau \cdot \beta \left( \frac{\theta_{i+1,j}^n - 2\,\theta_{i,j}^n + \theta_{i-1,j}^n}{(\Delta \xi)^2} + \frac{\theta_{i,j+1}^n - 2\, \theta_{i,j}^n + \theta_{i,j-1}^n}{(\Delta \eta)^2} \right)
\end{equation}

\noindent where \( \Delta \tau \) is the non-dimensional time step size and \( n \) denotes the time level.

The Dirichlet boundary conditions are applied on all four edges of the plate. The non-dimensional temperature values are specified as follows:

\begin{equation}
\theta(0, \eta, \tau) = \theta_{\text{left}}, \,\,\, \theta(1, \eta, \tau) = \theta_{\text{right}}, \,\,\, \theta(\xi, 0, \tau) = \theta_{\text{bottom}}, \,\,\, \theta(\xi, 1, \tau) = \theta_{\text{top}}.
\end{equation}

\noindent These conditions are implemented directly in the finite difference grid by setting the non-dimensional temperature values at the boundary nodes to the specified values at each time step. The initial non-dimensional temperature distribution \( \theta(\xi, \eta, 0) \) is specified for all interior nodes.

To ensure numerical stability, the time step \( \Delta \tau \) must satisfy the Courant-Friedrichs-Lewy (CFL) condition for the explicit Euler method,

\begin{equation}
\Delta \tau \leq \frac{1}{2\beta} \left( \frac{1}{(\Delta \xi)^2} + \frac{1}{(\Delta \eta)^2} \right)^{-1}
\end{equation}

The convergence of the solution is monitored by evaluating the change in the non-dimensional temperature distribution between successive time steps, continuing the simulation until the system reaches a sufficiently close approximation to a steady state.

Furthermore, in the challenge configurations the specification of the temperature on the boundaries includes additional features. In the Challenge-1 configuration, the left and right boundaries include segments of  predefined and fixed length, where the temperature is specified as follows:

\begin{equation}
\theta(\xi, \eta, \tau) = 
\begin{cases} 
1 & \text{for } \xi=0, \eta \text{ in the segment 1 range}, \\
0 & \text{for } \xi=1, \eta \text{ in the segment 2 range},\\
\text{randomly chosen in } [0, 1] & \text{for other boundary points.}
\end{cases}
\end{equation}

In the Challenge-2 configuration, the lengths of the boundary segments are again predefined and fixed, but their placement is randomized per simulation. Each segment spans a calculated portion of the grid points along the boundary, depending on the segment's length relative to the side's total length. If a segment is placed on the horizontal sides (\(\xi = 0\) or \(\xi = 1\)), it spans across the \(\eta\) dimension within a specified range; similarly, if on the vertical sides (\(\eta = 0\) or \(\eta = 1\)), it spans across the \(\xi\) dimension. The non-dimensional temperatures for the grid points within these segments are set to either 1 or 0, while the remainder of the boundary follows a specified pattern or random assignment within the normalized range, ensuring that segments do not overlap and appear on different sides:

\begin{equation}
\theta(\xi, \eta, \tau) = 
\begin{cases} 
1 & \text{for } \xi, \eta \text{ in the segment 1 range}, \\
0 & \text{for } \xi, \eta \text{ in the segment 2 range},\\
\text{randomly chosen in } [0, 1] & \text{for other boundary points.}
\end{cases}
\end{equation}
In the examples considered in this work, the length of the boundary segments was chosen to be four computational nodes, corresponding to ${1}/{6}$ of the interior nodes of the domain.

These configurations introduce an element of complexity and unpredictability in the thermal boundary conditions, challenging the model to adapt and learn from a wider array of scenarios than those it encountered during training. This setup is specifically designed to test the robustness of the model's generalization capabilities and its ability to discern and react to significant changes in boundary-driven thermal gradients. The impact of these boundary conditions on the learning process, particularly on the adaptation and emphasis of features by the final projection layer, is a focal point of analysis in this study. To illustrate this initialization, an example of two actual initial temperature fields used in the challenge configuration are shown below.

\begin{figure}[h!!]
    \centering
    \includegraphics[width=0.45\textwidth]{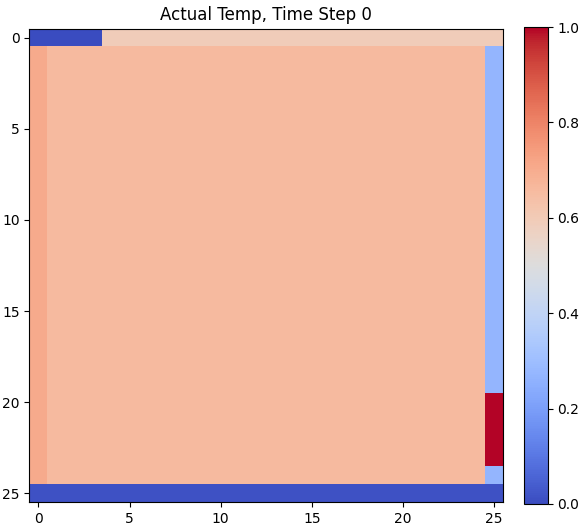}
    \includegraphics[width=0.45\textwidth]{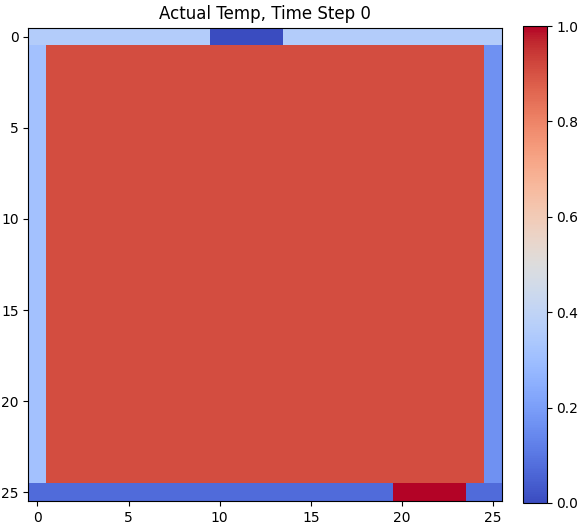}  
    \caption{Two initial temperature fields generated in the Challenge-2 configuration. The Dirichlet temperatures are chosen randomly in the range [0,1] (left, top and right) and [0,0.10] (bottom) independently of each other. The uniform internal temperature is also randomly initialized in [0, 1]. The two boundary segments are of equal and pre-specified length but their placement on different sides on the plate boundaries is random.  }
    \label{fig:domainCH2}
\end{figure}

\subsection{The physics-informed loss}
\label{sec:sub:PhysicsInformedLoss}

\newcommand{\thetaP}{{\theta_{\text{pred}}}}

An important aspect of the Transformer  model implementation is that the standard mean square error (MSE) loss is augmented with physics-informed components that we have found to be important in maintaining the alignment of the predictions with the underlying physics. In particular, the total loss \(L\) consists of the sum

\begin{equation}
L =  L_{\text{mse}} + \lambda_{\text{PI}} L_{\text{physics}} + \lambda_{\text{BC}} L_{\text{boundary}} + \lambda_{\text{IC}} L_{\text{initial}}
\label{eqn:Loss}
\end{equation}

\noindent where \(L_{\text{mse}}\) is the standard mean square error loss, \(L_{\text{physics}}\) is the physics-informed loss, \(L_{\text{boundary}}\) is the boundary condition loss, and \(L_{\text{initial}}\) is the initial condition loss. The weights \(\lambda_{\text{PI}}\), \(\lambda_{\text{BC}}\), and \(\lambda_{\text{IC}}\) balance the contributions of each component.

The physics-informed loss is derived by ensuring the model output aligns with the heat conduction equation. The residual is computed as:

\begin{equation}
\text{Residual} = \frac{\partial \thetaP}{\partial \tau} - \beta \left( \frac{\partial^2 \thetaP}{\partial \xi^2} + \frac{\partial^2 \thetaP}{\partial \eta^2} \right)
\end{equation}

The second-order spatial derivatives are approximated using central finite differences:

\begin{equation}
\frac{\partial^2 \thetaP}{\partial \xi^2} \approx \frac{\thetaP_{i+1,j} - 2\,\thetaP_{i,j} + \thetaP_{i-1,j}}{(\Delta \xi)^2}
\end{equation}

\begin{equation}
\frac{\partial^2 \theta}{\partial \eta^2} \approx \frac{\thetaP_{i,j+1} - 2\,\thetaP_{i,j} + \thetaP_{i,j-1}}{(\Delta \eta)^2}
\end{equation}

The temporal derivative is approximated as:

\begin{equation}
\frac{\partial \thetaP}{\partial \tau} \approx \frac{\thetaP_{i,j}^{n+1} - \thetaP_{i,j}^n}{\Delta \tau}
\end{equation}

The standard deviation of the residual is used to normalize the residuals:

\begin{equation}
\text{Residual}_{\text{std}} = \sqrt{\text{Var}(\text{Residual}) + \epsilon}
\end{equation}

\begin{equation}
\text{Normalized Residual} = \frac{\text{Residual}}{\text{Residual}_{\text{std}}}
\end{equation}

The physics-informed loss (\(L_{\text{physics}}\)) is then given by:

\begin{equation}
L_{\text{physics}} = \frac{1}{N} \sum_{i=1}^N \left( \text{Normalized Residual}_i \right)^2
\end{equation}
where \(N\) is the total number of residuals, calculated as the product of the number of samples, the number of grid points in the spatial domain (\(nx \times ny\)), and the number of time steps in the sequence.

The boundary condition loss (\(L_{\text{boundary}}\)) is calculated as the mean square error (MSE) between the predicted and expected Dirichlet values at the boundaries:

\begin{equation}
L_{\text{boundary}} = \text{MSE}(\theta_{\text{pred}}|_{\text{boundary}}, \theta_{\text{true}}|_{\text{boundary}})
\end{equation}

The initial condition loss (\(L_{\text{initial}}\)) is calculated as the MSE between the predicted and expected values at the initial (unmasked) time steps:

\begin{equation}
L_{\text{initial}} = \text{MSE}(\theta_{\text{pred}}|_{t=0}, \theta_{\text{true}}|_{t=0})
\end{equation}

The use of the physics-informed loss component in the model training scheme enforces alignment with the finite difference stencil for the heat conduction equation. This implicitly captures correlations between successive time steps by minimizing the residuals of the governing equations over the predicted sequence. Unlike in NLP applications, such as Large Language Models (LLMs), where a probability distribution over outcomes is predicted due to inherent uncertainty in language generation, our model predicts a deterministic solution for the temperature field. In probabilistic models, cross-correlation is useful for analyzing relationships between uncertain predictions. However, since our model produces a single outcome, temporal and spatial dependencies are already captured by the physics-based loss, making cross-correlation unnecessary. This approach is particularly relevant in block prediction mode, where outputs are predicted jointly and treated as independent, while in autoregressive mode, each prediction is conditioned on the previous ones, reducing the need to explicitly account for correlations between outputs.

\subsection{Implementation}

The numerical code is implemented in Python using the MLX framework and occasionally NumPy for array operations and is shared under an MIT license on GitHub\footnote{Repository available at \url{https://github.com/sck-at-ucy/MLX_BeyondLanguage}} \cite{mlxbeyond2024}.
MLX is used for core machine learning operations such as the definition of the model class for the physics-centric Transformer, the definition of loss functions and the training and validation process. Numpy is used for supporting utilities such as the generation of initial and boundary conditions and the creation of the datasets used for training, validation and testing. Conversion between MLX and Numpy arrays is trivial, e.g. 

\begin{figure}[h!!]
    \centering
\begin{center}
\begin{tcolorbox}[colback=gray!5!white, colframe=black, arc=0mm, outer arc=0mm, boxrule=0.1mm, width=0.75\textwidth, halign=left, boxsep=5mm]
\texttt{\textcolor{magenta!70!white}{import }\textcolor{black}{mlx.core }\textcolor{magenta!70!white}{as }\textcolor{black}{mx}}

\texttt{\textcolor{magenta!70!white}{import }\textcolor{black}{numpy }\textcolor{magenta!70!white}{as }\textcolor{black}{np}}

\texttt{\textcolor{magenta!70!white}{  }}

\texttt{\textcolor{black}{a }\textcolor{green!60!black}{= }\textcolor{black}{mx.arrange(}\textcolor{yellow!70!black}{3}\textcolor{black}{)}}

\texttt{\textcolor{black}{b }\textcolor{green!60!black}{= }\textcolor{black}{np.array(}\textcolor{yellow!70!black}{a}\textcolor{black}{)  }}\textcolor{yellow!40!black}{   \# copy of a}

\texttt{\textcolor{black}{c }\textcolor{green!60!black}{= }\textcolor{black}{mx.array(}\textcolor{yellow!70!black}{b}\textcolor{black}{)  }}\textcolor{yellow!40!black}{   \# copy of b}
\end{tcolorbox}
\end{center}
    \caption{Converting between MLX arrays and NumPy arrays.  }
    \label{fig:MLX_and_numpy}
\end{figure}

The code is structured to manage everything from initializing the geometrical setup and boundary conditions, through generating datasets, training, and evaluating the model, to plotting the results and saving the model for future use. Here's a breakdown of its main components and functionality:

\begin{itemize}
    \item \textbf{Random Seeds and Configuration:} The script starts by setting random seeds for reproducibility and defining configurations for model parameters, boundary conditions, and geometry of the 2D plate.
    \item \textbf{Geometry and Boundary Conditions:} It initializes the geometry and boundary conditions of the plate, deriving dimensions from the configuration and generating boundary conditions (BCs) and thermal diffusivities (alphas) for training, validation, and testing datasets.
    \item \textbf{Data Generation:} Uses a finite difference method to generate datasets for the thermal simulation of the plate. This involves setting up a grid based on the plate dimensions and iterating over time steps to simulate heat distribution according to the given BCs and thermal diffusivities.
    \item \textbf{Model Definition:} A masked Transformer  model is implemented, specifically designed to handle the spatial and temporal aspects of the heat diffusion problem. It includes positional encodings for different dimensions and a Transformer  encoder architecture tailored to this application.
    \item \textbf{Model Predictions Modes:} The model can be trained either in a Block Prediction mode or an Autoregressive Stepwise Prediction mode. The basic structure of the model remains largely unmodified and what enables the two alternative modes is the use of different attention masks.
    \item \textbf{Training and Validation:} The training process involves forward passes to compute losses including a physics-informed loss that integrates knowledge of the heat equation. Normalization of the various loss components is key for achieving best model performance. Updates of the model's weights are done via the Adam optimizer. Validation occurs concurrently to gauge the model's performance on unseen data. The code leverages the MLX compile function transformation, which compiles computation graphs. Function compilation results in smaller graphs by merging common work and fusing certain operations and in our case achieves approximately 25\% speedup during training.
    \item \textbf{Testing the Trained Model:} Testing of the trained model takes into account whether the model has been trained to operate as a Block Predictor or an Autoregressive Stepwise Predictor. The testing loss reported in each case reflects this.
    \item \textbf{Plotting and Saving:} It can plot the model's predictions against the actual data to visually assess performance. It also provides functionality to save both the model's parameters and the configuration settings to ensure experiments are easily reproducible.
    \item \textbf{Utility Functions:} Several utility functions support the main processes, including generating the initial datasets, loading data in batches, calculating derivatives for the physics-informed loss, and more.
\end{itemize}

The code is structured for development and experimentation, indicating ongoing adjustments and optimizations, such as tuning the physics-informed loss and expanding the model's extrapolation capabilities. It's a comprehensive approach to integrating machine learning with physical simulations, aiming to leverage the strengths of Transformers  for spatial-temporal data. In this sense, this article is meant to offer a starting point for anybody wanting to explore the potential of Transformers  in engineering physics and the associated code should be understood as work in progress. Even so, we have made an effort to debug and test thoroughly before sharing. The main focus of the discussion that follows is on explaining the structure of the Transformer  model.

\subsection{The Transformer Model Class}
\label{section:sub:Model}

We define a custom, physics-centric Transformer model class that is designed specifically to handle the 2D structure of the data. The actual code and  associated model parameters are given in \ref{sec:appendixA}.

\begin{figure}[h!!]
    \centering
    \includegraphics[width=0.9\textwidth]{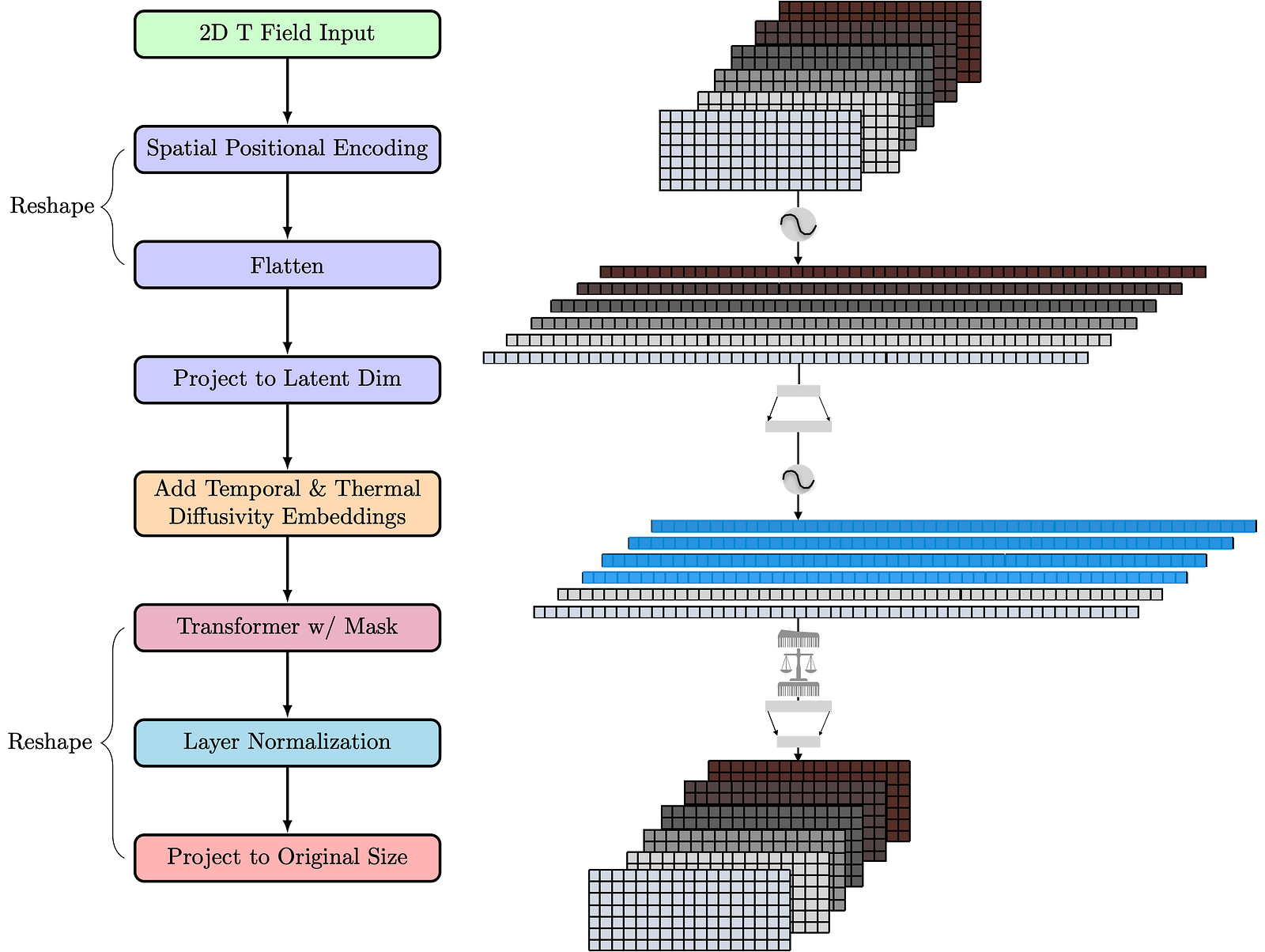}

    \caption{Custom physics-centric Transformer model (shown in block prediction mode.)}
    \label{fig:flowchart}
\end{figure}

The Transformer uses spatial sinusoidal positional encodings and temporal sinusoidal positional encoding which are embedded in the latent dimension along with a thermal diffusivity embedding.
Given the specific nature of the heat conduction problem at hand, we opted to omit the decoder component of the Transformer architecture. While the traditional encoder-decoder framework is crucial for sequence-to-sequence tasks, such as those common in NLP, it offers no significant advantage in our case. Here, the goal is to predict the entire temperature evolution of a 2D plate from given boundary and initial conditions — tasks that are sufficiently handled by the self-attention mechanism of the encoder alone. Including the decoder would introduce unnecessary complexity, adding more layers and parameters without yielding meaningful improvements in accuracy or performance. We tested both configurations, and the decoder’s contribution was marginal. Therefore, to maintain an efficient model with a lean architecture suitable for personal machines, we opted for an encoder-only approach, which is fully capable of capturing the long-term spatial and temporal dependencies required to predict temperature distributions accurately. The exact sequence of events in the \textcolor{red!65!blue}{\_\_call\_\_} method is summarized in the diagram of  figure~\ref{fig:flowchart}.

It is easier to handle the flow of information when the data is flattened in a 1D form. However, it is also easier to generate the spatial positional embeddings before the input data is flattened. Thus, it is worth stopping for a second to take a closer look at the sequence of operations. The spatial positional embeddings are generated and added before flattening the input data. Then the spatially-embedded data is flattened and projected to the latent dimension through a linear layer. Then the temporal positional and heat diffusivity embeddings\footnote{While the use of spatial and temporal embeddings aligns directly with standard practice in NLP, it is worth explaining the embedding of the heat diffusivity, which varies from case to case but has no spatial or temporal variation. Essentially, this represents a transformation that maps the diffusivity into a new representation space that the model can utilize alongside other embeddings.} are added to the latent dimension of the flattened data. The masked Transformer  processes the data and returns its normalized prediction, which is then projected back to the original shape through another linear projection layer.

Furthermore,  the Transformer  is masked. This means that even though the model receives the entire ground truth sequence as input, it has access only to selected frames. The inclusion of the attention mask allows to operate in different predictive modes with minimal modifications. The intended predictive mode determines which input frames are visible to the model.  
We differentiate between two alternative model uses and their respective trainings strategies, Block Predictions and Autoregressive Stepwise Predictions. In Block Prediction mode, the intention is to train the model to predict entire evolution histories. Given a small number of initial steps in the evolution of the temperature field, the model learns to output the entire history to steady state at once. The Block Prediction mode is a powerful approach for scenarios where you can define clear initial states and expect the model to generate coherent future sequences based on those states. This mode is particularly useful in initial value problems in fields like physics (those where initial conditions can significantly shape system evolution). Other possible uses include finance for predictive simulations based on observed trends, or even creative fields like music and art generation where an initial motif can inspire a complete piece, but those lie beyond our interest here.
In the Autoregressive Stepwise mode, the model is trained to predict the next frame in the evolution or at most a small number of forward frames (steps). In this scenario, the model could potentially be used to march a solution forward by sampling its own previous predictions. This mode aligns more directly with the standard uses of Transformers.
From a coding perspective, the challenge is to define a model class and training strategy that can accommodate both modes. Here, we leverage the Transformer's attention mask to achieve both predictive modes.  Thus in both cases the basic model architecture is largely the same, what differs is the attention mask used.  Irrespective of the predictive mode, during training the Transformer  receives batches of complete time evolution histories of the entire 2D temperature field. That is, the data passed to the model is contained in a tensor (MLX array) of the shape

\begin{center}
\begin{tcolorbox}[colback=gray!5!white, colframe=black, arc=0mm, outer arc=0mm, boxrule=0.1mm, width=0.65\textwidth, halign=center, boxsep=5mm]
\texttt{\textcolor{magenta!70!white}{src}\textcolor{black}{[}\textcolor{blue}{batch\_size}\textcolor{black}{, }\textcolor{blue}{seq\_len}\textcolor{black}{, }\textcolor{blue}{nx}\textcolor{black}{, }\textcolor{blue}{ny}\textcolor{black}{]}}
\end{tcolorbox}
\end{center}

\noindent where \textcolor{blue}{batch\_size} is the number of datasets in each batch, \textcolor{blue}{seq\_len} is the number of time steps in the time evolution history, and (\textcolor{blue}{nx}, \textcolor{blue}{ny}) is the size of the discretization grid. Its predictions are returned in the same shape. Our Transformer model predicts the full 
$nx \times ny$ spatial grid at each time step, with masking applied only along the temporal dimension. The entire spatial frame is predicted simultaneously for each time step, allowing the model to capture both spatial and temporal dependencies. While the temporal sequence is masked to ensure autoregressive or block prediction behavior, the spatial dimensions are processed jointly at each step.

In MLX, as in other machine learning frameworks, elements of the attentions mask that are set to a very large negative value (provided by MLX as  \textcolor{blue}{-inf}) hide the 
corresponding input sequence elements from the Transformer. Where the mask is set to zero, the corresponding input sequence is unmasked.  

\begin{figure}[h!!]
    \centering
    \begin{subfigure}[b]{\textwidth}
        \centering
        \includegraphics[width=1.0\textwidth]{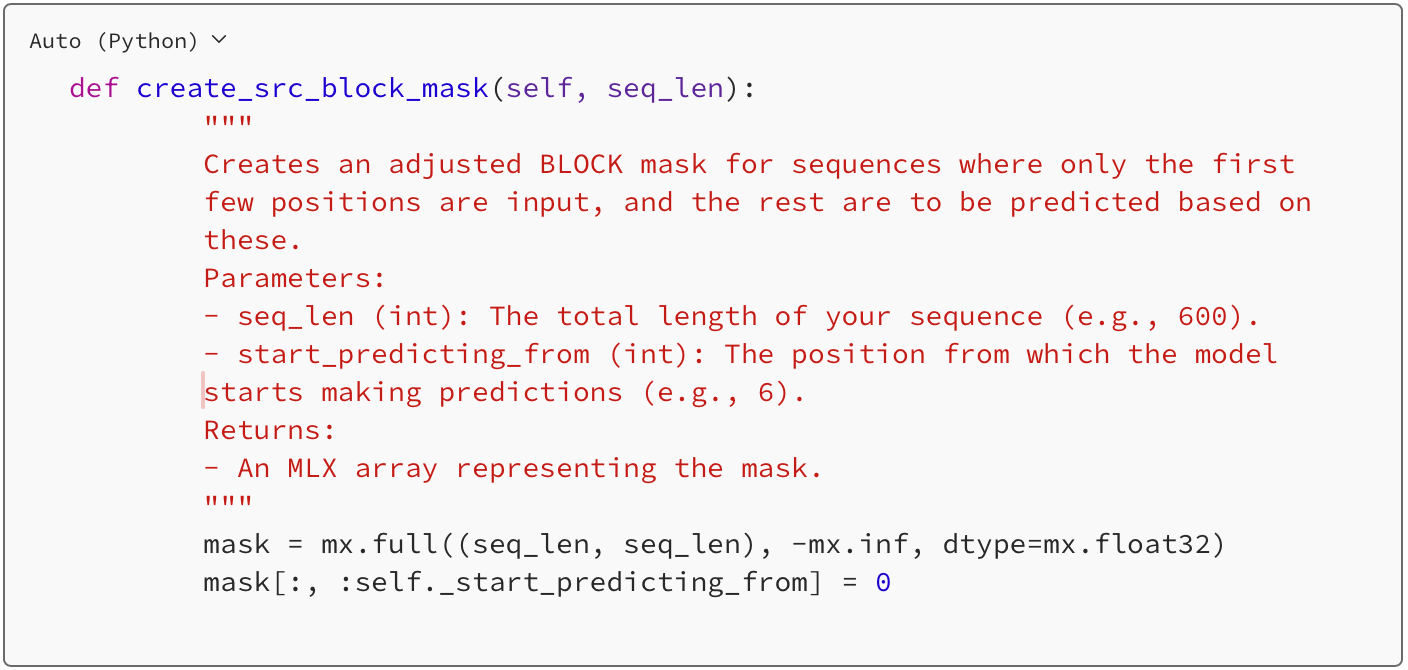}
        \caption{}
        \label{fig:MLX_block_mask}
    \end{subfigure}
    \hfill
    \begin{subfigure}[b]{\textwidth}
        \centering
        \includegraphics[width=1.0\textwidth]{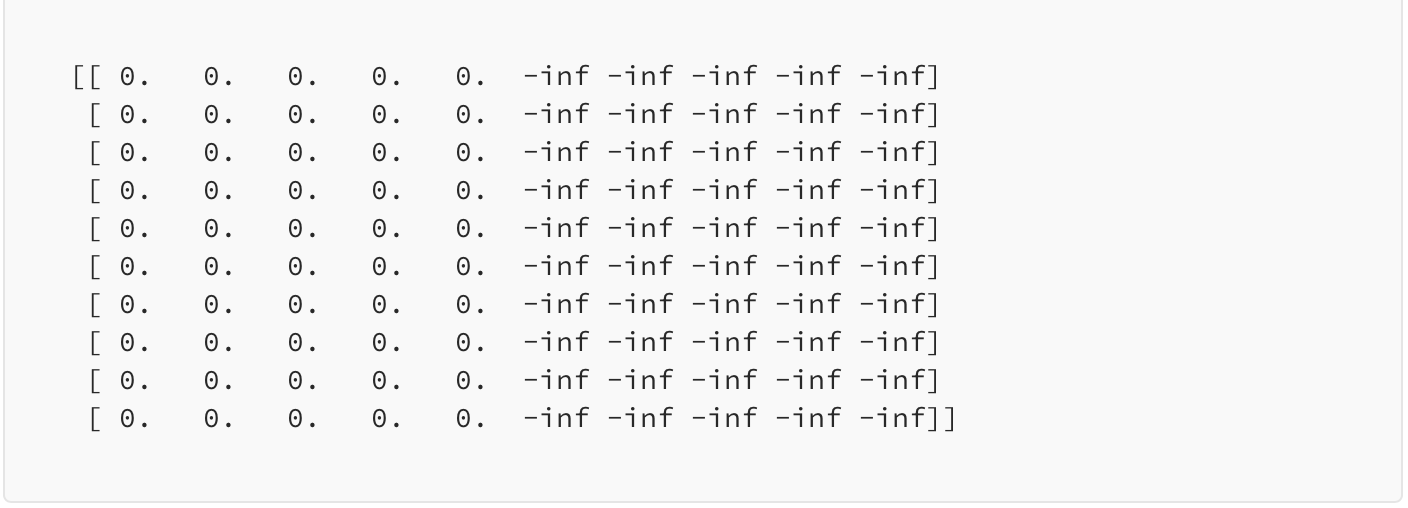}
        \caption{}
        \label{fig:block_mask}
    \end{subfigure}
    \caption{(a) MLX code for generating the block mask.; (b) Example of a block mask used in the Block Prediction mode, assuming   \textcolor{blue}{seq\_len} = 10 and 5 unmasked initial frames.}
    \label{fig:code_and_mask_block}
\end{figure}

In the case of the Block Prediction mode, the mask has all the columns set to  \textcolor{blue}{-inf} except for the first $N\_input$ that are set to 0. These columns correspond to the $N\_input$  initial frames that are accessible to the model and used for the prediction of the entire subsequent sequence block. The code defining the Block attention mask is fairly simple (see Figure~\ref{fig:MLX_block_mask}).  To help visualize the Block attention mask, Figure~\ref{fig:block_mask} shows how it would look for the hypothetical scenario where \textcolor{blue}{seq\_len} = 10 and 5 unmasked initial frames are provided as input.

\begin{figure}[h!!]
    \centering
    \begin{subfigure}[b]{\textwidth}
        \centering
        \includegraphics[width=1.0\textwidth]{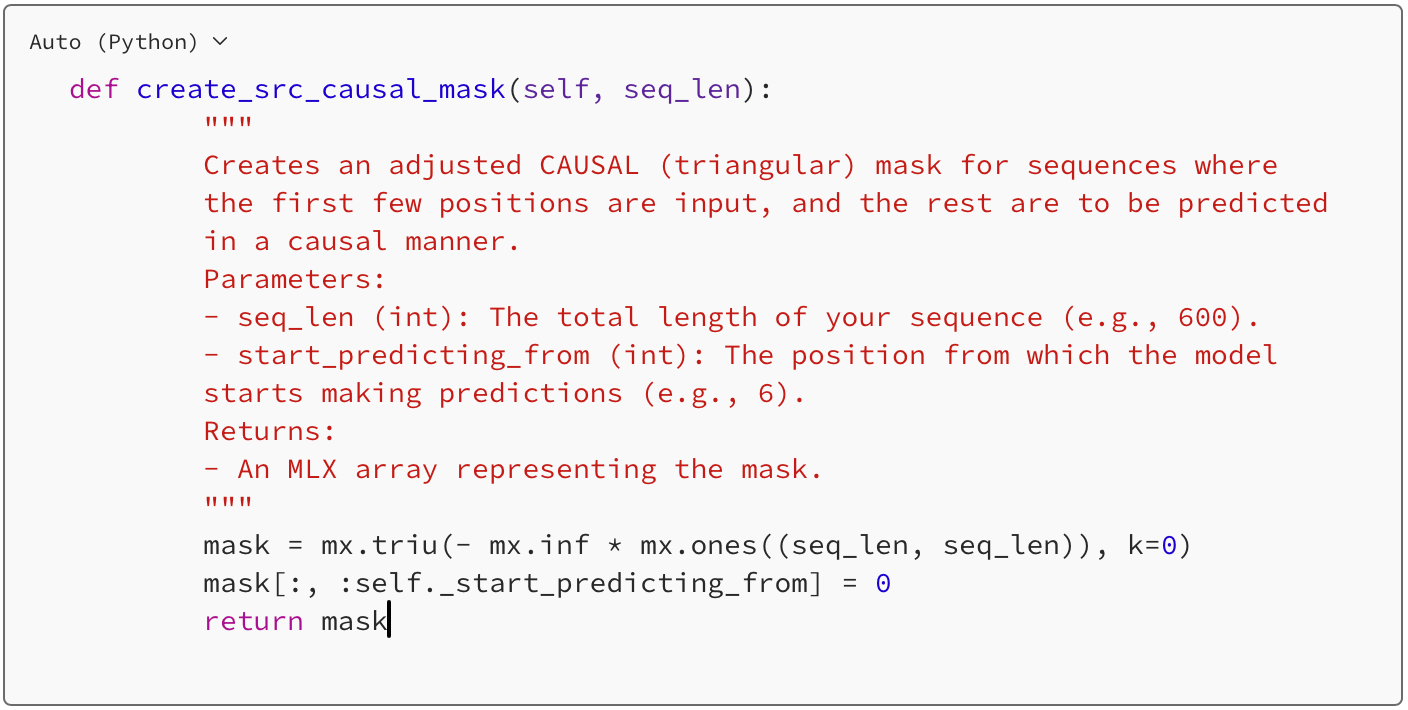}
        \caption{}
        \label{fig:CausalMaskCode}
    \end{subfigure}
    \hfill
    \begin{subfigure}[b]{\textwidth}
        \centering
        \includegraphics[width=1.0\textwidth]{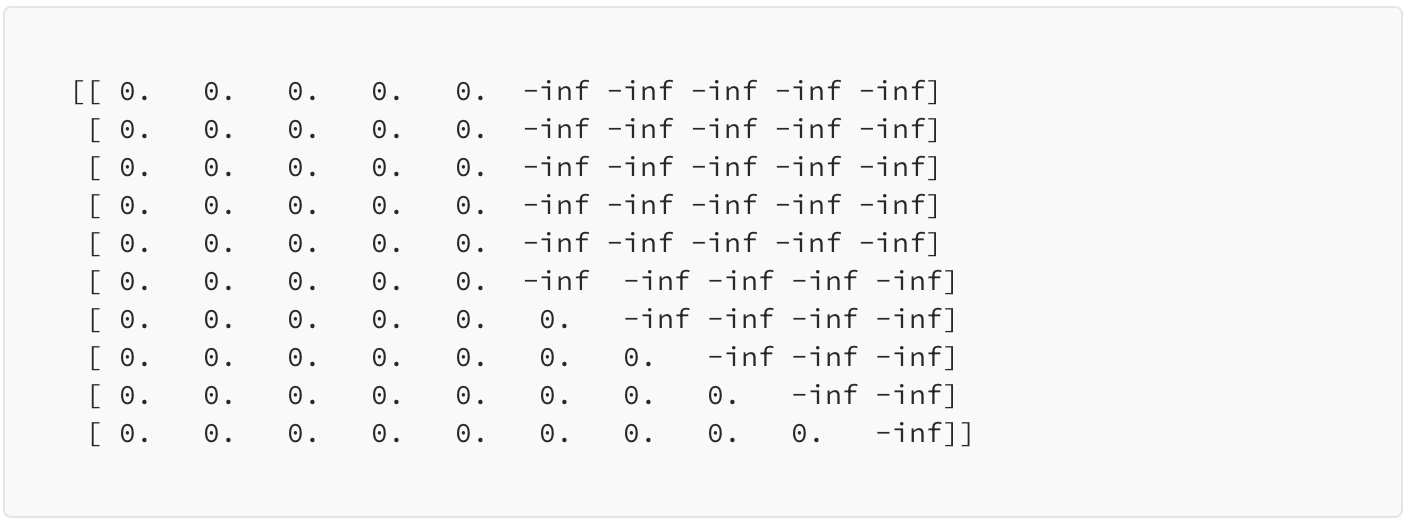}
        \caption{}
        \label{fig:CausalMaskExample}
    \end{subfigure}
    \caption{(a) MLX code for generating the causal mask.; (b) Example of a causal mask used in the Autoregressive mode, assuming   \textcolor{blue}{seq\_len} = 10 and 5 unmasked initial frames.}
    \label{fig:code_and_mask_causal}
\end{figure}

 \noindent Contrast this to the Triangular Causal Mask that is used in the case of Autoregressive Stepwise mode (see Figure~\ref{fig:CausalMaskCode}). In this case, we again allow a few initial frames to be visible but then switch to a triangular causal mask that prevents the model from peeking into the future during training. Figure~\ref{fig:CausalMaskExample} depicts how the Causal attention mask for the Autoregressive mode would look like, again assuming the hypothetical scenario of  \textcolor{blue}{seq\_len} = 10 with 5 unmasked initial frames provided as input.

Thus, in predicting any particular frame the model can access the initial five frames plus (if applicable) all frames up to but excluding the frame itself. You might wonder, if the mask's diagonal should also be zero in the predictive (triangular) region, as that is often the case in the application of Transformers to LLM. However, here we eventually want to use the trained model to march the solution to steady state starting from a few known initial frames in the sequence. Allowing access to the frame being predicted during training would murk the true predictive capacity of the model.

\subsection{Training, Validation and Testing Strategy}
\label{section:sub"Training}
The training and validation datasets are divided into batches. The same batch size is used both for training and validation. During training, for each field in a batch, the entire time history of the 2D temperature distribution is passed to the Transformer. Because the Transformer  is masked it is only allowed to "see" (access) the frames specified by either the Block or Autoregressive Causal mask.
The training is done over a user-specified number of epochs. For each epoch, we compute a training and a validation loss. The loss has several components as already discussed in equation~(\ref{eqn:Loss}). 

The Transformer  predictions and the associated losses are computed for all points and all times at once. Given sufficient memory to accommodate the problem size, the process is highly efficient, especially given the lazy evaluation of the compute graphs that MLX follows.

Assessing the true inference performance of the trained Transformer must be approached differently in the Block and Autoregressive Stepwise modes. In the case of Block predictions, the model learns to predict the entire sequence to steady state all at once. Thus, when the trained Block Prediction model is applied to the test field, the loss is computed as the MSE between the ground truth sequence and the model's entire output sequence (in our case the MSE loss is augmented with physics-based losses as done during training). On the other hand, the application of the trained Autoregressive Stepwise model to the test field is more involved. The model is fed the entire ground truth sequence but the causal mask ensures strict sequential predictions. Then, we enter an iteration cycle, where the model prediction is repeatedly sampled to replace one more frame at a time in the model input, starting from the first frame after the unmasked initial frames (note: there could be a single or a handful of unmasked initial frames). Each modified input results in a new model prediction and this iteration is repeated till the entire ground truth input sequence is replaced with model predictions. In a sense, the model is used to march the solution forward without being aware of the future ground truth values beyond the frame that is being replaced at each iteration stage. In this case, the loss is computed as the MSE loss between the iteratively (autoregressively) generated sequence and the ground truth and not between the initial model output and the ground truth. This is a more demanding assessment of the model performance since the error can progressively accumulate as the solution is marched forward and more and more frames in the input sequence are replaced by predictions. With a poorly performing model, the solution could in fact diverge significantly from the ground truth.

\section{Results}

\label{sec:results}

\subsection{Results for the Base Case (block prediction mode)}
\label{sec:sub:ResultsBaseBlock}

\begin{figure}[h!!]
    \centering
    \includegraphics[width=0.9\textwidth]{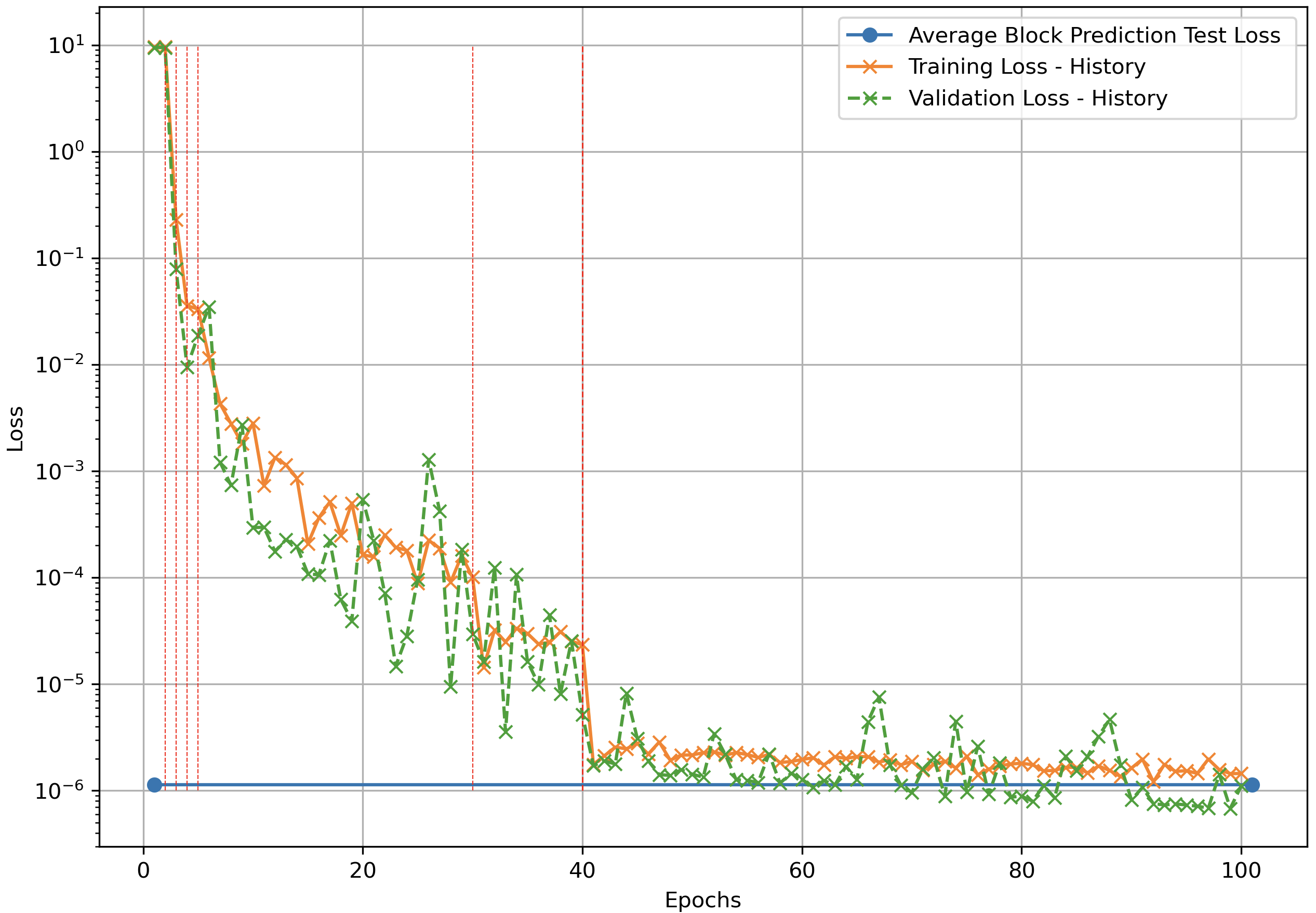}
    \caption{Evolution of the training and validation losses using a learning rate schedule for the block prediction mode. Learning rate transitions are marked with red dashed lines.  }
    \label{fig:CH1TraininValidationLosses_BASE_BLOCK}
\end{figure}

The model was trained for 100 epochs in batches of 4 with a training set of 8400 sets (70$\%$) and an in-line validation using 2400 sets (20$\%$). The completely hidden test set comprised of 1200 sets (10$\%$). The evolution of the training and validation loss are shown in Figure~\ref{fig:CH1TraininValidationLosses_BASE_BLOCK}.  An MLX learning rate scheduler was implemented and the learning rate transition points are marked by red vertical dashed lines. The schedule can be summarized as follows, 

 \begin{equation}
LR  = 
\begin{cases} 
 0,                              &\text{epoch: } 1,  \text{ warmup},  \\
 1\mathrm{e}{-5},      &\text{epoch: } 2,   \text{ initial phase ramp-up}, \\
 1\mathrm{e}{-4},      &\text{epoch: } 3,   \text{initial phase ramp-up}, \\
 5\mathrm{e}{-4},      &\text{epoch: } 4,   \text{initial phase ramp-up},\\
 1\mathrm{e}{-3},      &\text{epochs: } 5 \text{ to } 30,   \text{initial phase},\\
 5\mathrm{e}{-4},      &\text{epochs: } 31\text{ to } 40,   \text{ramp-down for physics-informed,}\\
1\mathrm{e}{-4},       &\text{epochs: } 41\text{ to } 100,  \text{physics-informed loss dominated.}\\
\end{cases}
\label{eqn:baseschedule}
\end{equation}

\begin{figure}[h!]
    \centering
    \includegraphics[width=\textwidth]{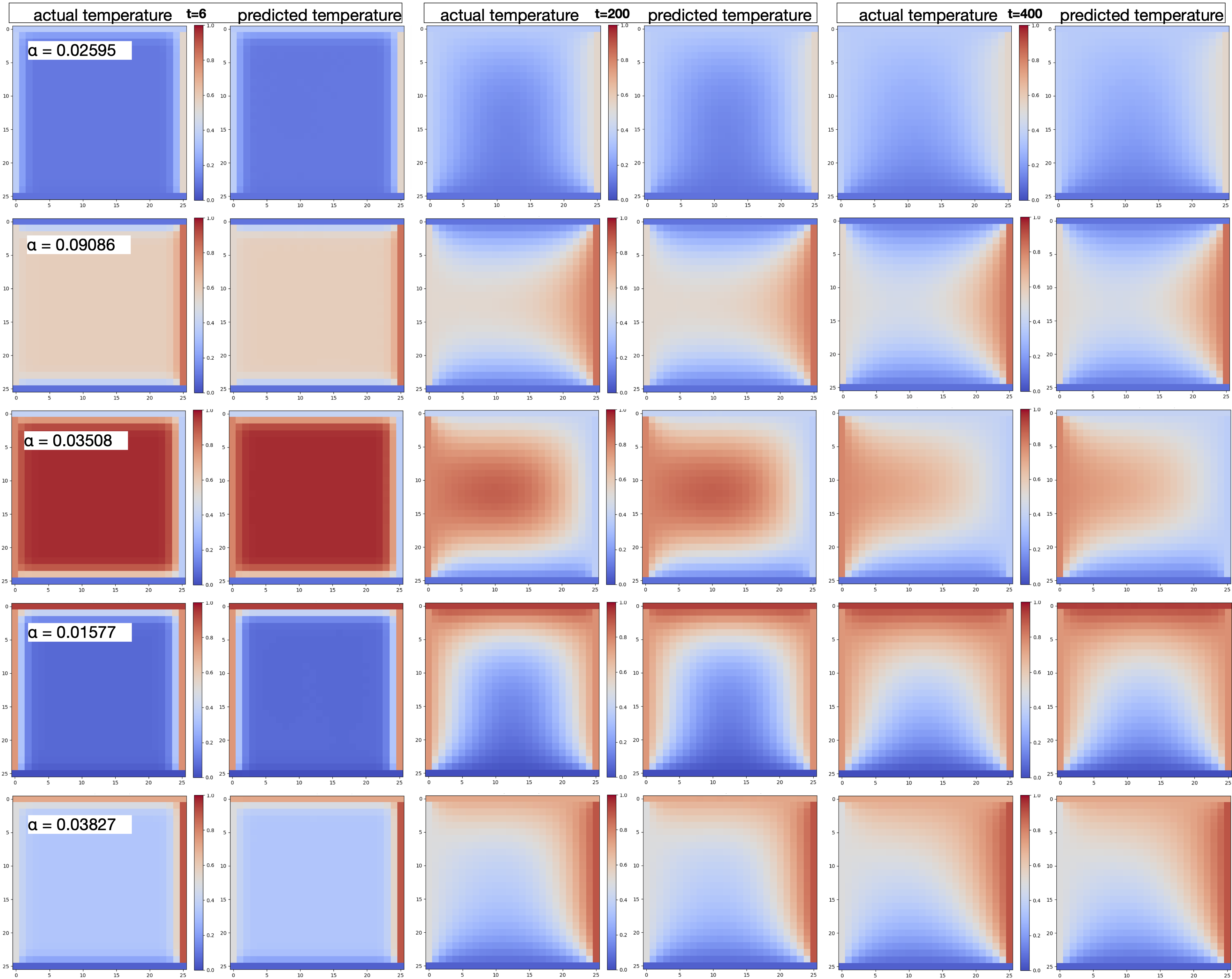}
    \caption{Model inference performance in block prediction mode for the base case. Each row corresponds to an unknown test case with unique boundary and initial conditions and unique thermal diffusivity. In each row, the first two frames show the comparison of the ground truth to the model predictions at dimensionless time t=6 (first masked instance), the middle two frames at time t=200 and the right two frames show the state state at time t=400. }
    \label{fig:Fames_Base_Block}
\end{figure}

The average loss over the 1200 cases of the test set is  1.13e-06. This value represents the average over all sets. The loss for each test case is computed as the average over the entire evolution history from the initial condition to the final steady state frame.  A visual impression of the model inference performance is shown in Figure~\ref{fig:Fames_Base_Block} for four randomly selected test cases. In each case, a comparison is made between the ground truth (left panel) and the model prediction (right panel), for dimensionless times t=6 (first masked instance), t=200 (mid-way) and t=400 (final steady state). No differences can be identified, confirming the exceedingly low value of test loss.

\eject

\subsection{Results for the Base Case (auto-regressive prediction mode)}
\label{sec:sub:ResultsBaseRegressive}

The training for the model for autoregressive predictions was again carried out over 100 epochs in batches of 4 using the same learning rate schedule as described in equation~(\ref{eqn:baseschedule}).
\begin{figure}[ht]
    \centering
    \includegraphics[width=0.9\textwidth]{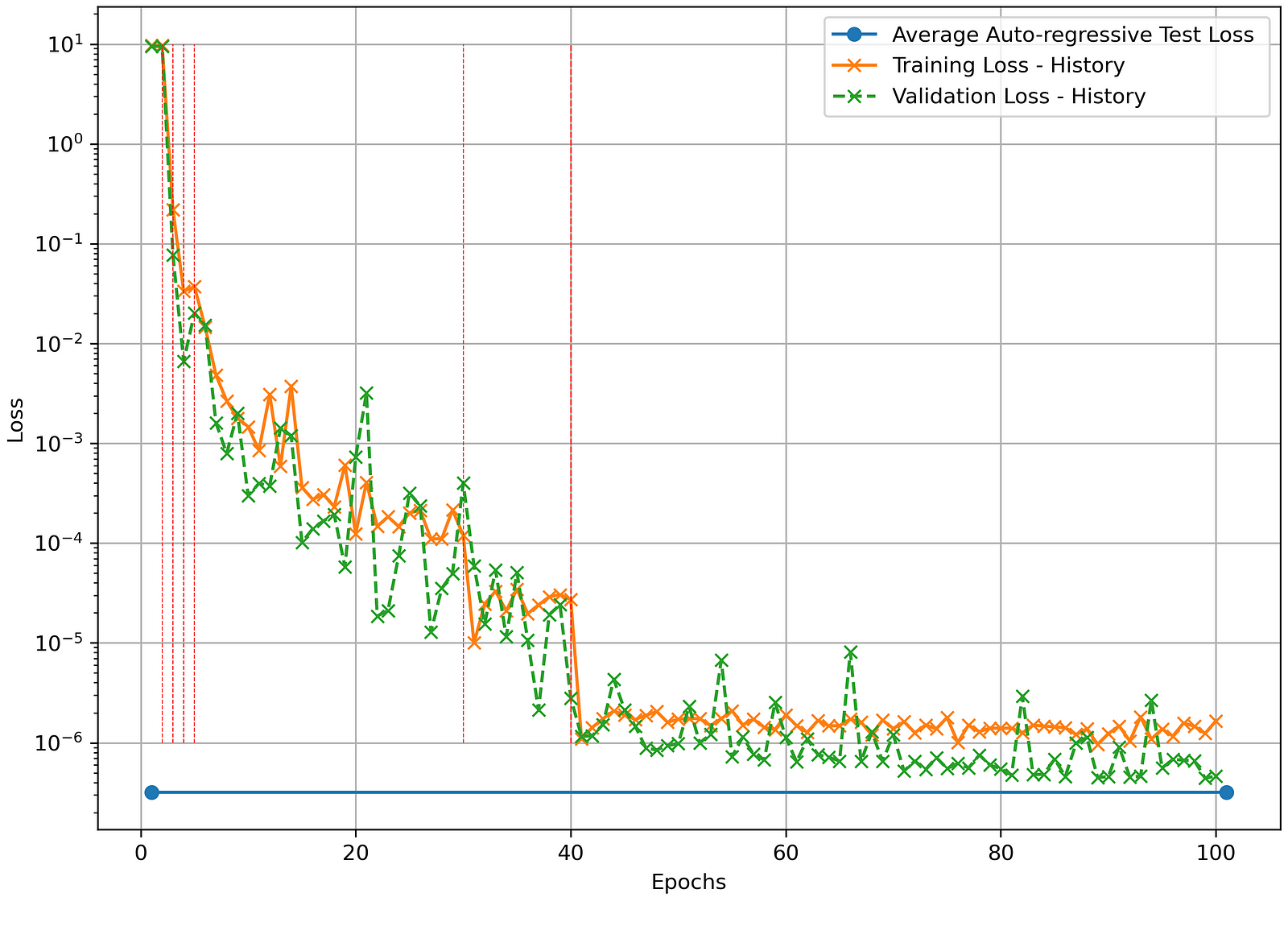}

    \caption{Evolution of the training and validation losses using a learning rate schedule for the autoregressive mode. Learning rate transitions are marked with red dashed lines.  }
    \label{fig:CH1TraininValidationLosses_BASE_REGRESSIVE}
\end{figure}
As can be seen in Figure~\ref{fig:CH1TraininValidationLosses_BASE_REGRESSIVE}, the model learns effectively and both the training and validation losses decrease steadily by more than seven orders of magnitude. The average autoregressive test loss of 3.2e-07 was computed over all 1200 hidden test cases. For each individual test case, the loss was the computed as the average over all regressive steps.  The fact that the test loss is well-aligned (in fact slightly lower than the training and validation losses) suggests that the model is able to handle successfully previously unseen combinations of parameters (i.e. boundary conditions, initial internal distribution and heat diffusivity). 

\begin{figure}[h!!]
    \centering
    \includegraphics[width=\textwidth]{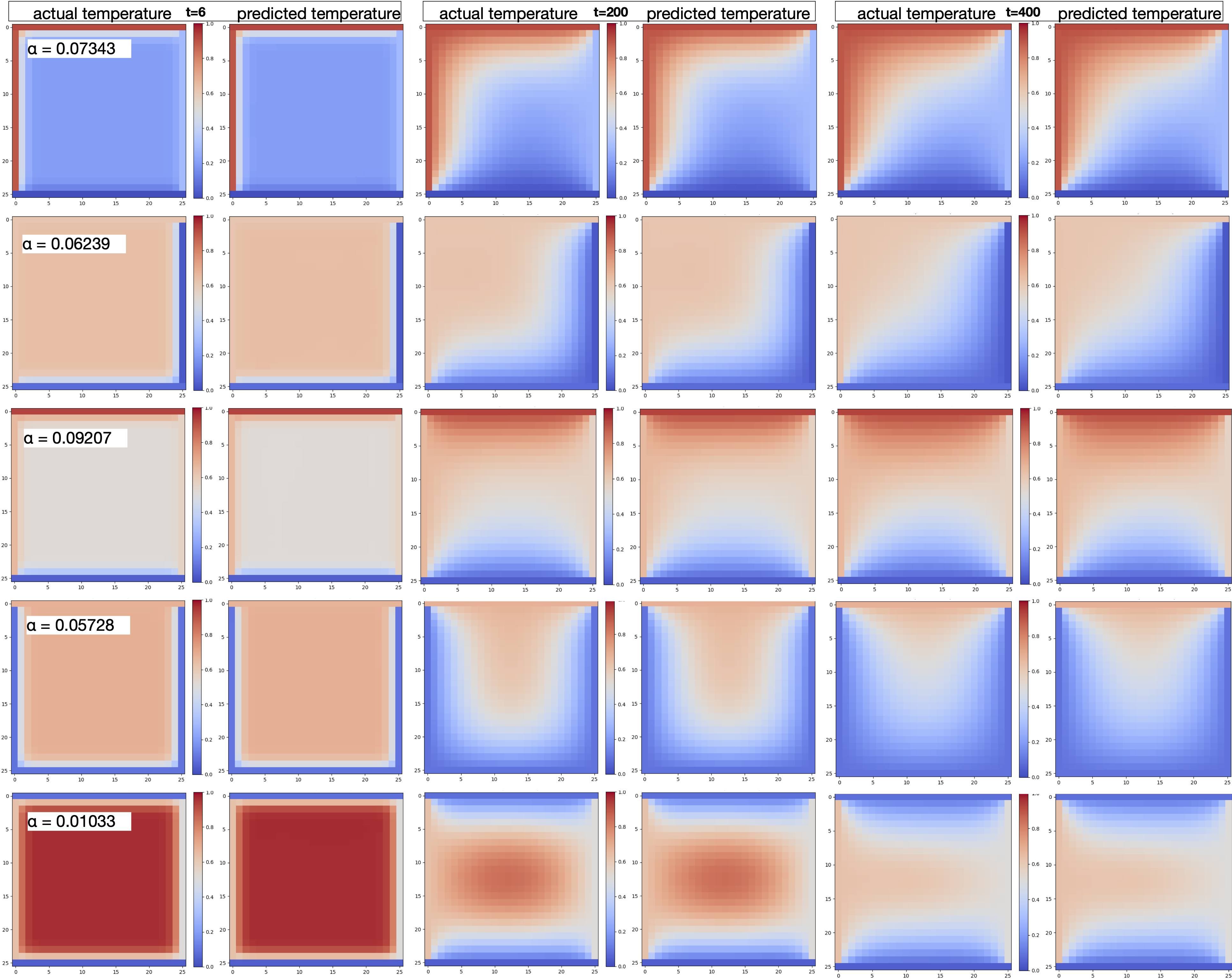}
    \caption{Model inference performance in autoregressive prediction mode for the base case. Each row corresponds to an unknown test case with unique  boundary and initial conditions and  unique thermal diffusivity. In each row, the first two frames show the comparison of the ground truth to the model predictions  at dimensionless time t=6, the middle two frames at time t=200 and the right two frames show the state state at time t=400. }
    \label{fig:Challenge1_regressive}
\end{figure}

 The good model inference performance is also reflected in the visual comparisons of Figure~\ref{fig:Challenge1_regressive}, where frames of the ground-truth and regressive model predictions are shown side-by-side for dimensionless times of t=6, 200 and 400. Each row corresponds to a randomly selected test case. 

\eject

\break


\subsection{Results for Challenge 1 (block prediction mode)}
\label{sec:sub:ResultsChallenge1block}

\begin{figure}[h!!]
    \centering
    \includegraphics[width=0.9\textwidth]{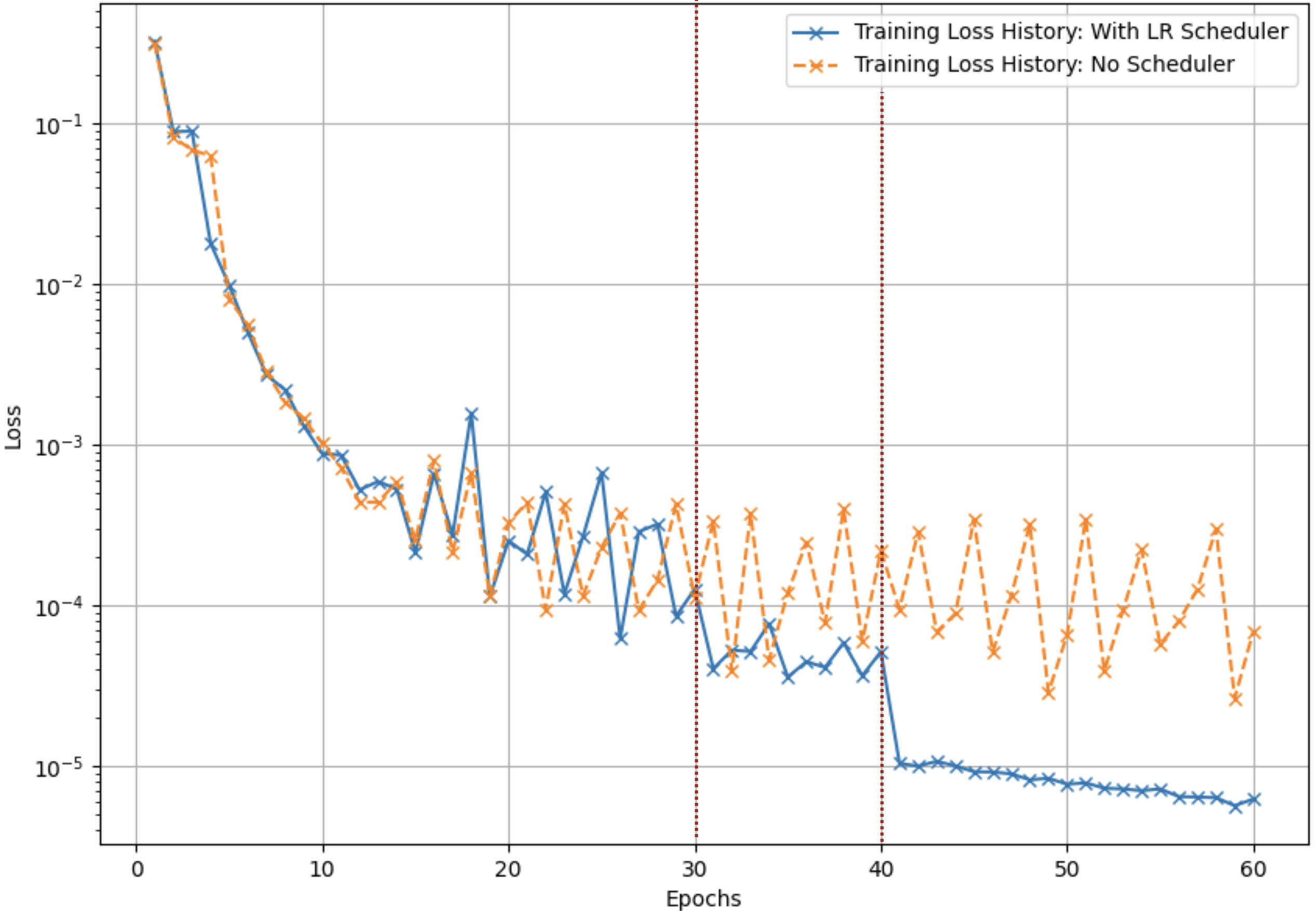}
    \caption{The time history of the of the training loss with and without the use of a learning rate scheduler.}
    \label{fig:CH1LearningRateScheduler}
\end{figure}

Challenge-1 introduces additional localized variabilities in the evolution of the temperature field and offers a good example to highlight the importance of using a learning rate schedule. We have found that at later stages of the training, the total loss comes to be dominated by the physics-informed component that enforces awareness of the underlying physical laws. While this itrend is true even in the base case, in Challenge-1 the physics-informed loss becomes more important. During this later stage, a lower learning rate is essential to enable the model to continue learning effectively. To illustrate the importance of using a learning rate schedule, especially for the more complex configurations of Challenge 1 and Challenge 2, Figure~\ref{fig:CH1LearningRateScheduler} shows the evolution of the total loss with and without a schedule in the case of Challenge 1. As a result of these observations for Challenge-1, we have adopted a learning rate schedule in all cases (Base, Challenge-1, Challenge-2) for consistency.

\begin{figure}[t!!]
    \centering
    \includegraphics[width=0.9\textwidth]{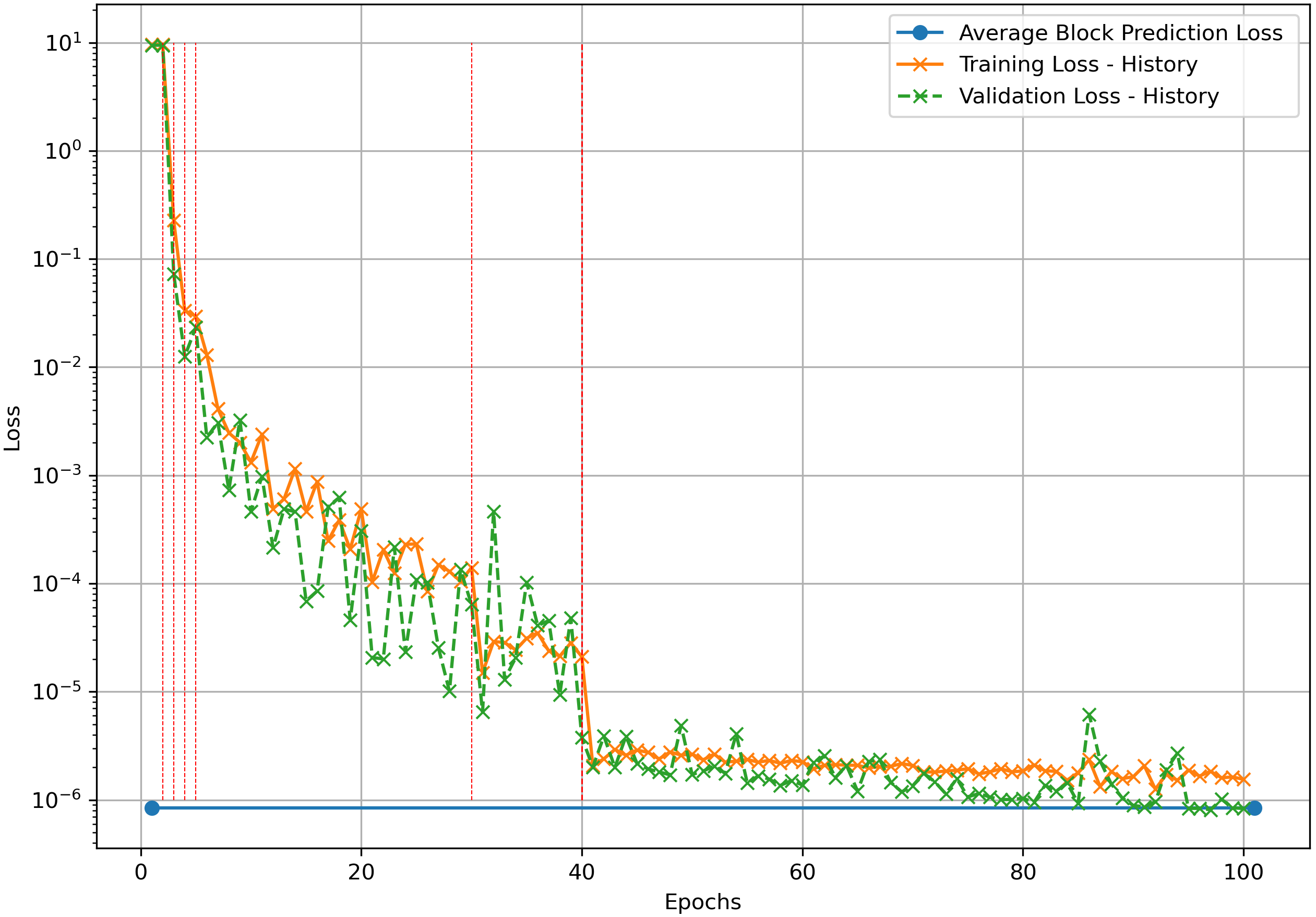}

    \caption{ Evolution of the training and validation losses using a learning rate schedule with transitions indicated by red dashed lines. The average test loss for inference in block prediction mode  is indicated by the solid blue line.}
    \label{fig:CH1TraininValidationLosses_CH1_Block}
\end{figure}

Following the same strategy as for the Base case, the model was trained for 100 epochs in batches of 4 with a training set of 8400 sets (70$\%$)and an in-line validation using 2400 sets (20$\%$). The completely hidden test set comprised of 1200 sets (10$\%$). The evolution of the training and validation loss are shown in Figure~\ref{fig:CH1TraininValidationLosses_CH1_Block}.  The learning rate schedule of equation~(\ref{eqn:baseschedule}) was used  and the learning rate transition points are marked by red vertical dashed lines.

\begin{figure}[h!!]
    \centering
    \includegraphics[width=\textwidth]{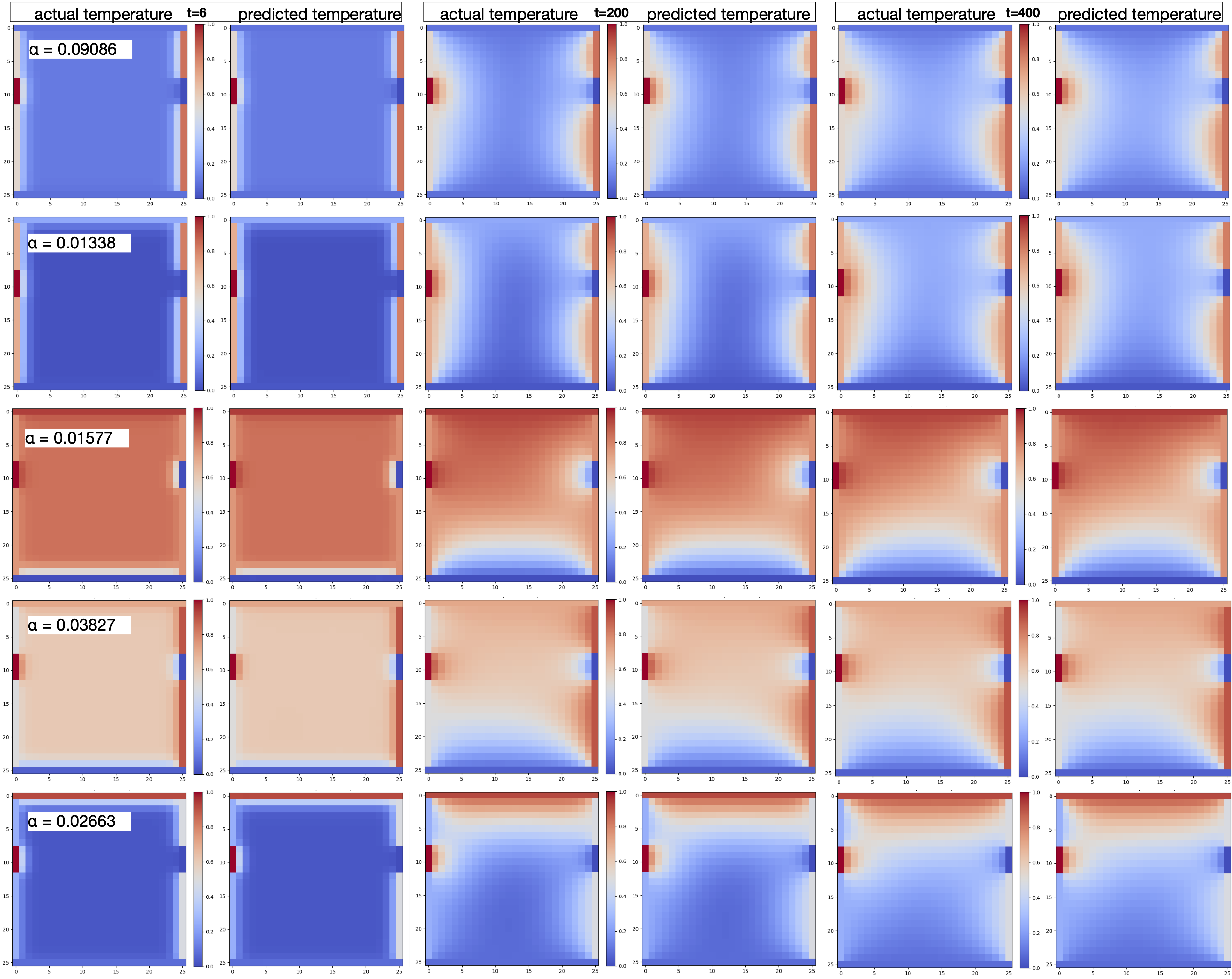}

     \caption{Same as Fig.~\ref{fig:Fames_Base_Block} but  for inference in block prediction mode for Challenge-1.} 
    \label{fig:Challenge1_block}
\end{figure}

The average loss over the 1200 cases of the test set is  8.43e-07. This value represents the average over all sets. The loss for each test case is computed as the average over the entire evolution history from the initial condition to the final steady state frame.   A visual impression of the model performance is shown in Figure~\ref{fig:Challenge1_block}
where model predictions and the ground truth are shown side-by-side for the initial (t=6), midway (t=200) and final frames (t=400) for several randomly  chosen cases from the test set. 
There are no obvious deviations between the ground truth and the predicted frames. Note, that the model is able to capture the localized temperature gradients associated with the fixed boundary segments, irrespective of changes in other features, i.e. Dirichlet boundary conditions, initial internal temperature distribution and thermal diffusivity. Video animations of the temperature evolution histories for several test cases can be found in the supplemental material posted on GitHub\cite{mlxbeyond2024}.  

\eject


\subsection{Results for Challenge 1 (auto-regressive prediction mode)}
\label{sec:sub:ResultsChallenge1regressive}

For the autoregressive mode of Challenge-1, the model was trained using the same learning rate schedule, batch size and total number of epochs as for the block prediction mode. What is impressive is that the average test loss for autoregressive inference is slightly lower than the final training and validation losses achieved during training. The reason is that during training the loss is computed on the fly over the entire sequence of frames to avoid spending an excessive time in auto-regression over the large training set. 

\begin{figure}[t!!]
    \centering
    \includegraphics[width=0.9\textwidth]{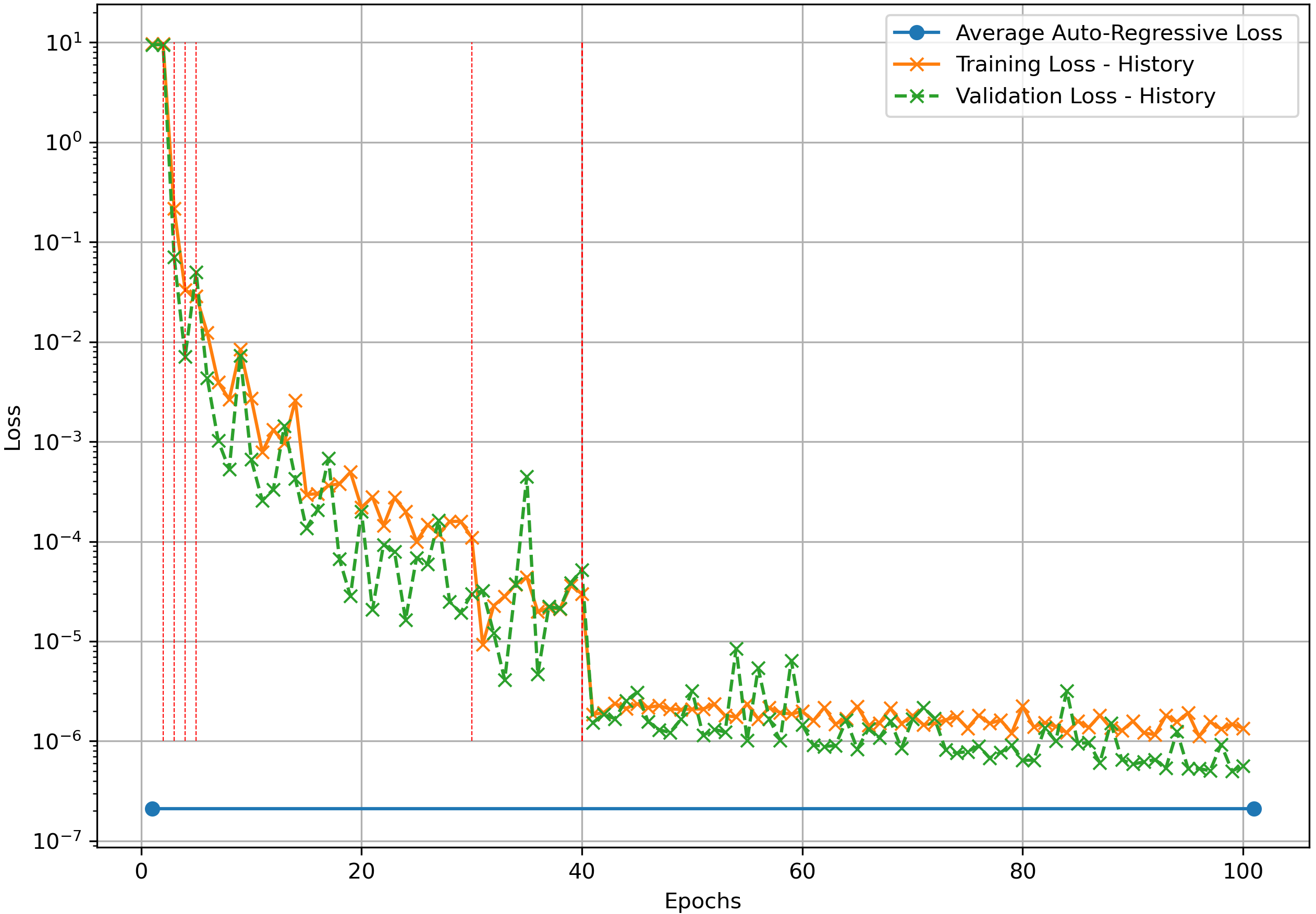}
    \caption{ Evolution of the training and validation losses using a learning rate schedule with transitions indicated by red dashed lines. The average test loss for inference in autoregressive  mode  is indicated by the solid blue line.}
    \label{fig:CH1TraininValidationLosses_CH1_Reg}
\end{figure}

\begin{figure}[h!!]
    \centering
    \includegraphics[width=\textwidth]{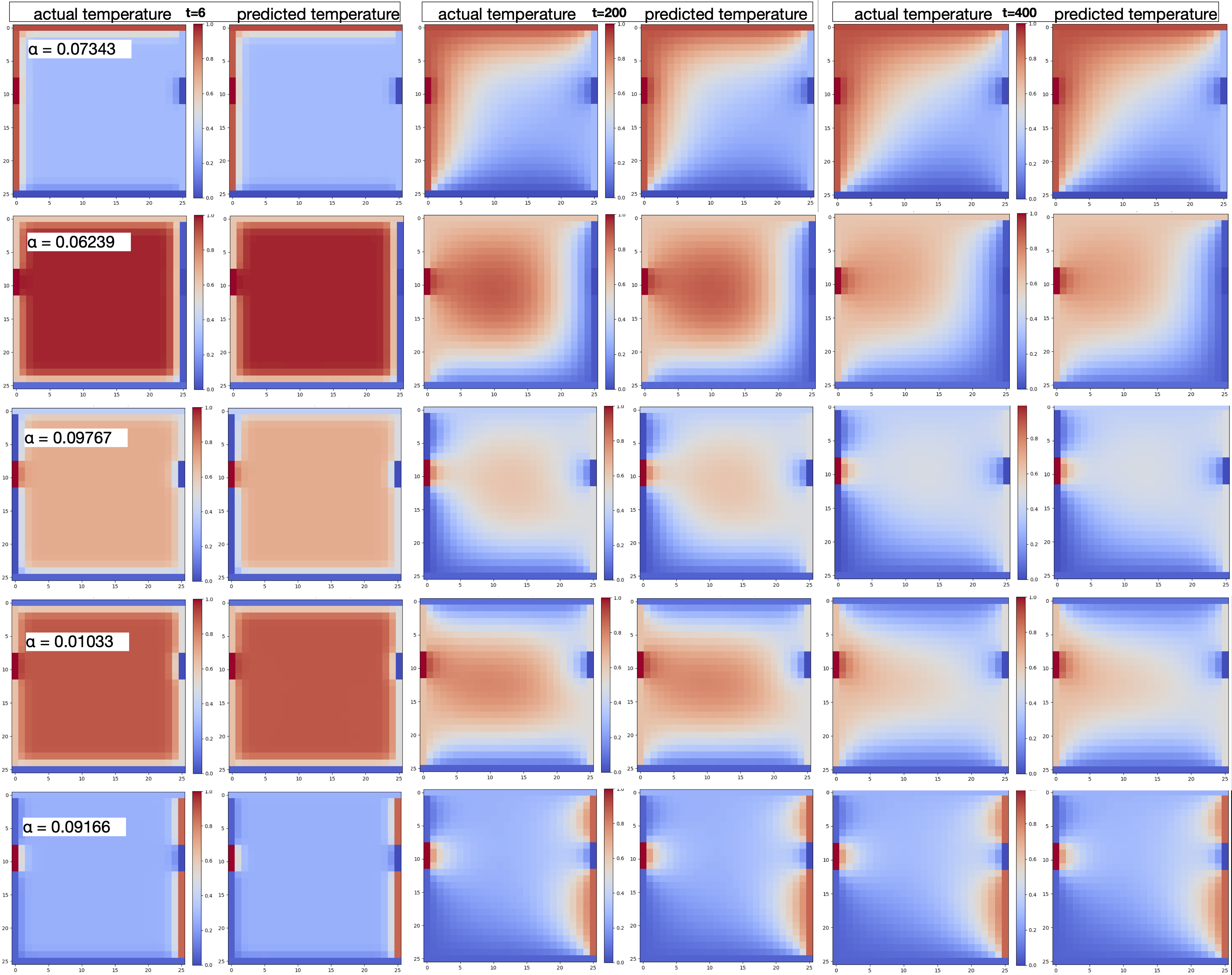}

     \caption{Same as Fig.~\ref{fig:Fames_Base_Block} but  for inference in auto-regressive mode for Challenge-1.} 
        \label{fig:Challenge1_autoregressive}
\end{figure}

The comparison of the ground truth and model autoregressive prediction at selected times is shown in Figure~\ref{fig:Challenge1_autoregressive} following the same format as for the block prediction case. The model predictions are indistinguishable from the ground truth, which is remarkable given that the model is used to march forward the solution to steady state by repeatedly sampling its own predictions.

\subsection{Results for Challenge 2 (block prediction mode)}
\label{sec:sub:ResultsChallenge2block}

We have found that the enhanced range of variability introduced by the random boundary segments in Challenge-2 made necessary some modifications to the  model and its training to preserve the same level of model performance. The number of encoder layers in Transformer was increased from 12 to 24 (see~\ref{sec:appendixA}). The total number of epochs was increased to 300 and the learning schedule was adjusted accordingly as shown in equation~(\ref{eqn:CH2schedule}).

 \begin{equation}
LR  = 
\begin{cases} 
 0,                              &\text{epoch: } 1,  \text{ warmup},  \\
 1\mathrm{e}{-5},      &\text{epoch: } 2,   \text{ initial phase ramp-up}, \\
 1\mathrm{e}{-4},      &\text{epoch: } 3,   \text{initial phase ramp-up}, \\
 5\mathrm{e}{-4},      &\text{epoch: } 4,   \text{initial phase}, \\
 1\mathrm{e}{-4},      &\text{epochs: } 41\text{ to } 100,   \text{ramp-down for physics-informed,}\\
5\mathrm{e}{-5},       &\text{epochs: } 101\text{ to } 200,  \text{physics-informed loss dominated,}\\
1\mathrm{e}{-5},       &\text{epochs: } 201\text{ to } 300,  \text{physics-informed loss dominated}\\
\end{cases}
\label{eqn:CH2schedule} 
\end{equation}

\begin{figure}[h!!]
    \centering
    \includegraphics[width=0.9\textwidth]{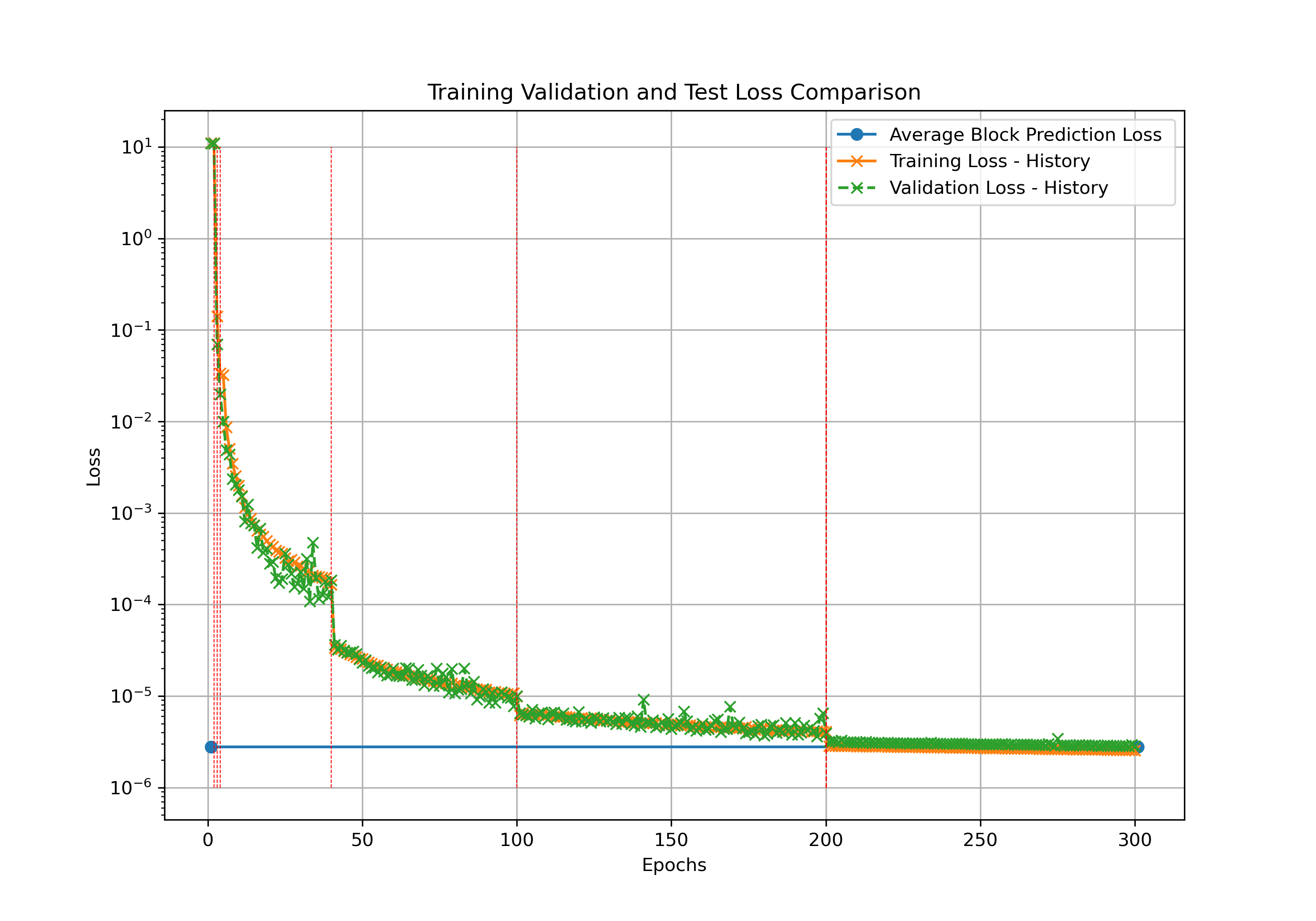}
    \caption{ Evolution of the training and validation losses using the learning rate schedule of equation~(\ref{eqn:CH2schedule}) for Challenge-2 with transitions indicated by red dashed lines. The average test loss for inference in block prediction  mode is  indicated by the solid blue line.}
    \label{fig:CH2TraininingBlockValidationLosses}
\end{figure}

The evolution history of the training and validation losses based on this schedule are shown in Figure~\ref{fig:CH2TraininingBlockValidationLosses}.  The average test loss for block prediction inference has a value of  is 2.78e-06 as indicated by the solid blue line. The validation loss tracks closely the training loss and the model learns effectively. Nevertheless, the final loss values achieved are moderately higher than in the base and Challenge-1 cases, which is not surprising considering the complexity of Challenge-2. The comparison of the ground truth and model predictions using frames at dimensionless times t=6, 200 and 400 confirm the inference performance of the model when challenged with unknown cases in the test set. The model is able to capture well the regions of localized gradients in the vicinity of the random boundary segments.  

\begin{figure}[h!!]
    \centering
    \includegraphics[width=\textwidth]{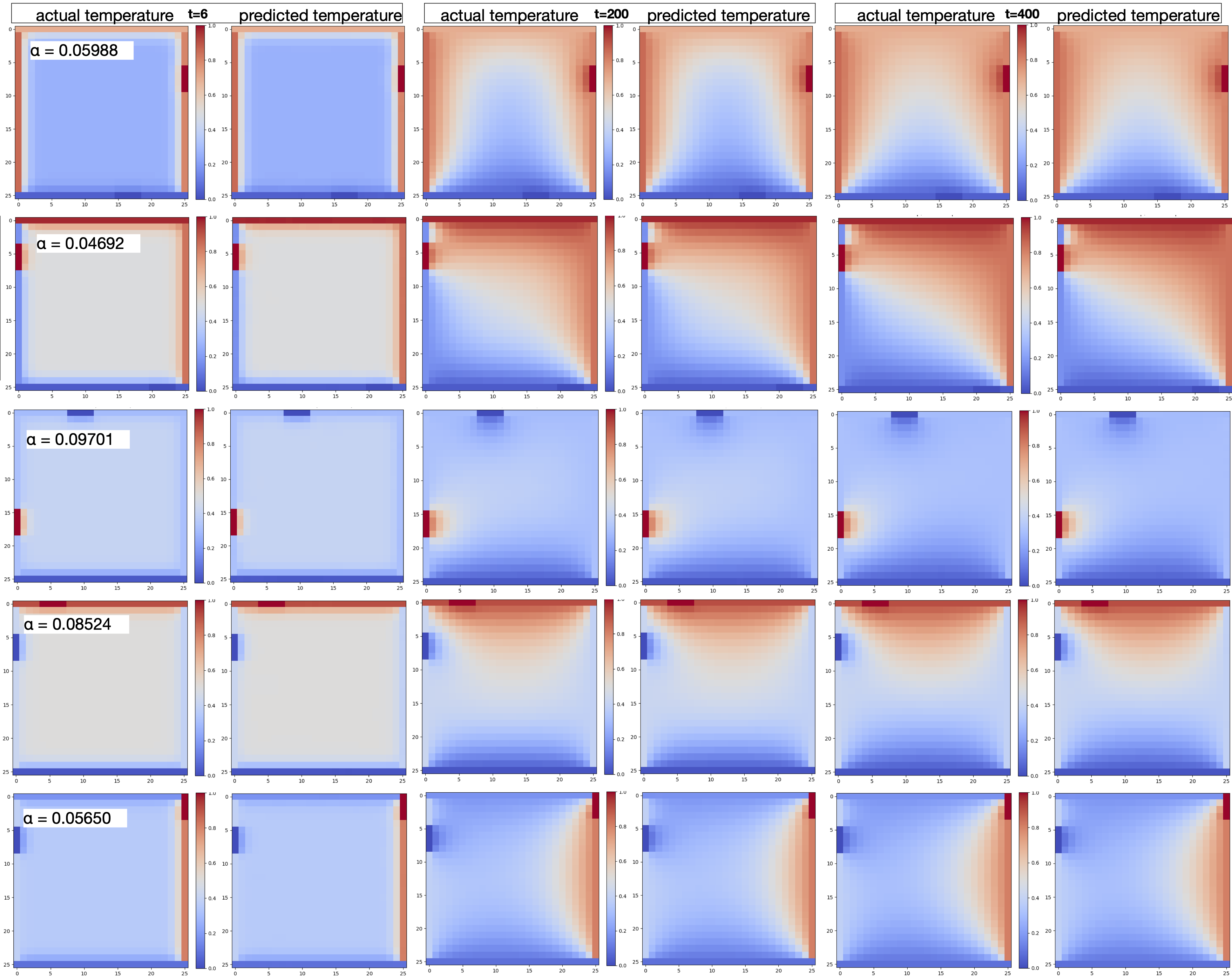}
        \caption{Same as Fig.~\ref{fig:Fames_Base_Block} but  for inference in block prediction mode in Challenge-2. }
    \label{fig:Challenge2_block}
\end{figure}

\break

\subsection{Results for Challenge 2 (auto-regressive prediction mode)}
\label{sec:sub:ResultsChallenge2regressive}

\begin{figure}[h!!]
    \centering
    \includegraphics[width=0.9\textwidth]{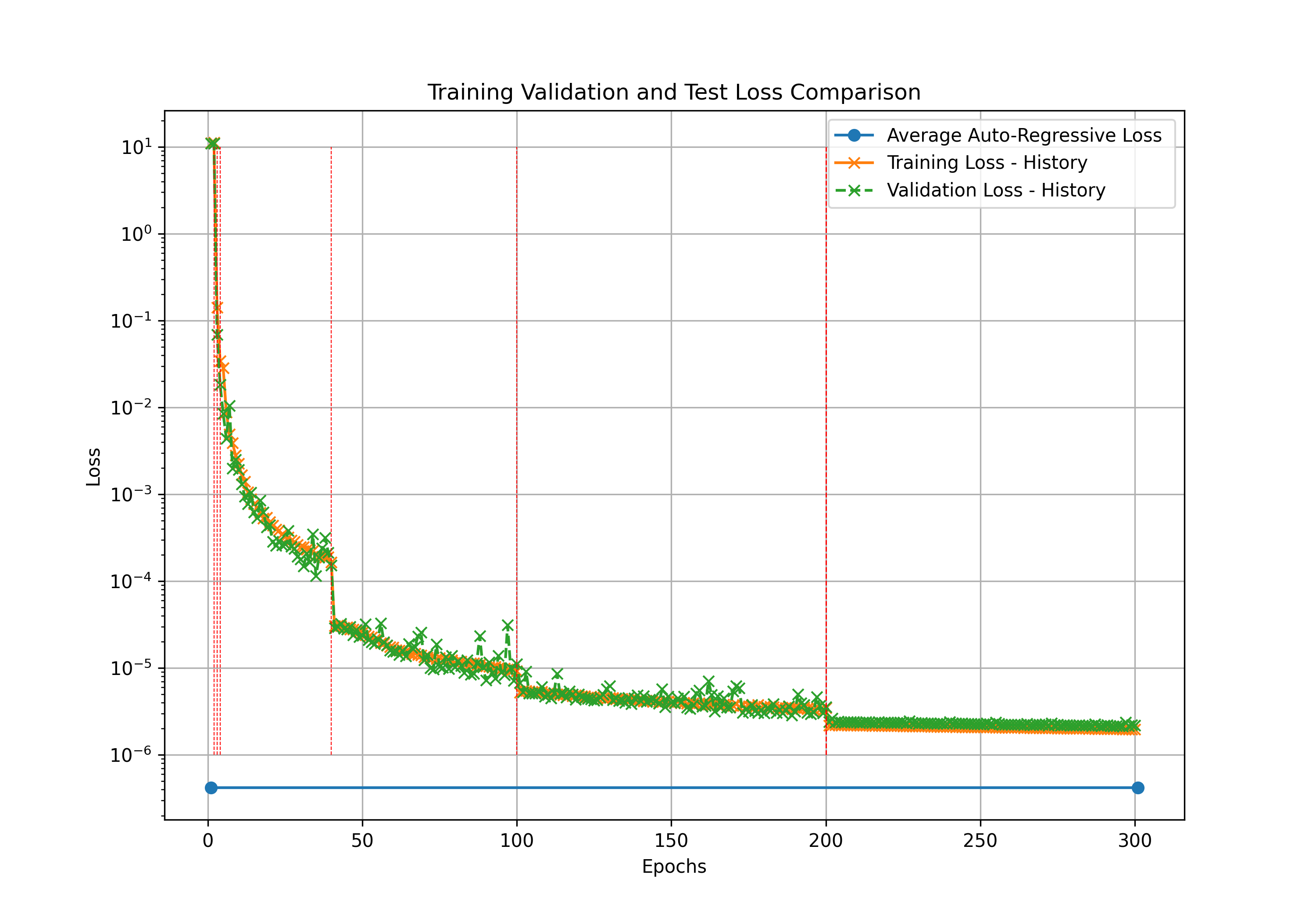}
    \caption{Evolution of the training and validation losses using the learning rate schedule of equation~(\ref{eqn:CH2schedule}) for Challenge-2 with transitions indicated by red dashed lines.  The average test loss for inference in autoregressive prediction  mode is  indicated by the solid blue line.}
    \label{fig:CH2TrainingRegressValidationLosses}
\end{figure}

As shown in Figure~\ref{fig:CH2TrainingRegressValidationLosses}, using the same learning rate schedule of equation~(\ref{eqn:CH2schedule}) for training the model in the autoregressive mode yields a similar evolution history for the training and validation losses. The average test loss for inference in the autoregressive mode is indicate by the blue solid line and corresponds to a value of 4.21e-07. 

\begin{figure}[h!!]
    \centering
    \includegraphics[width=\textwidth]{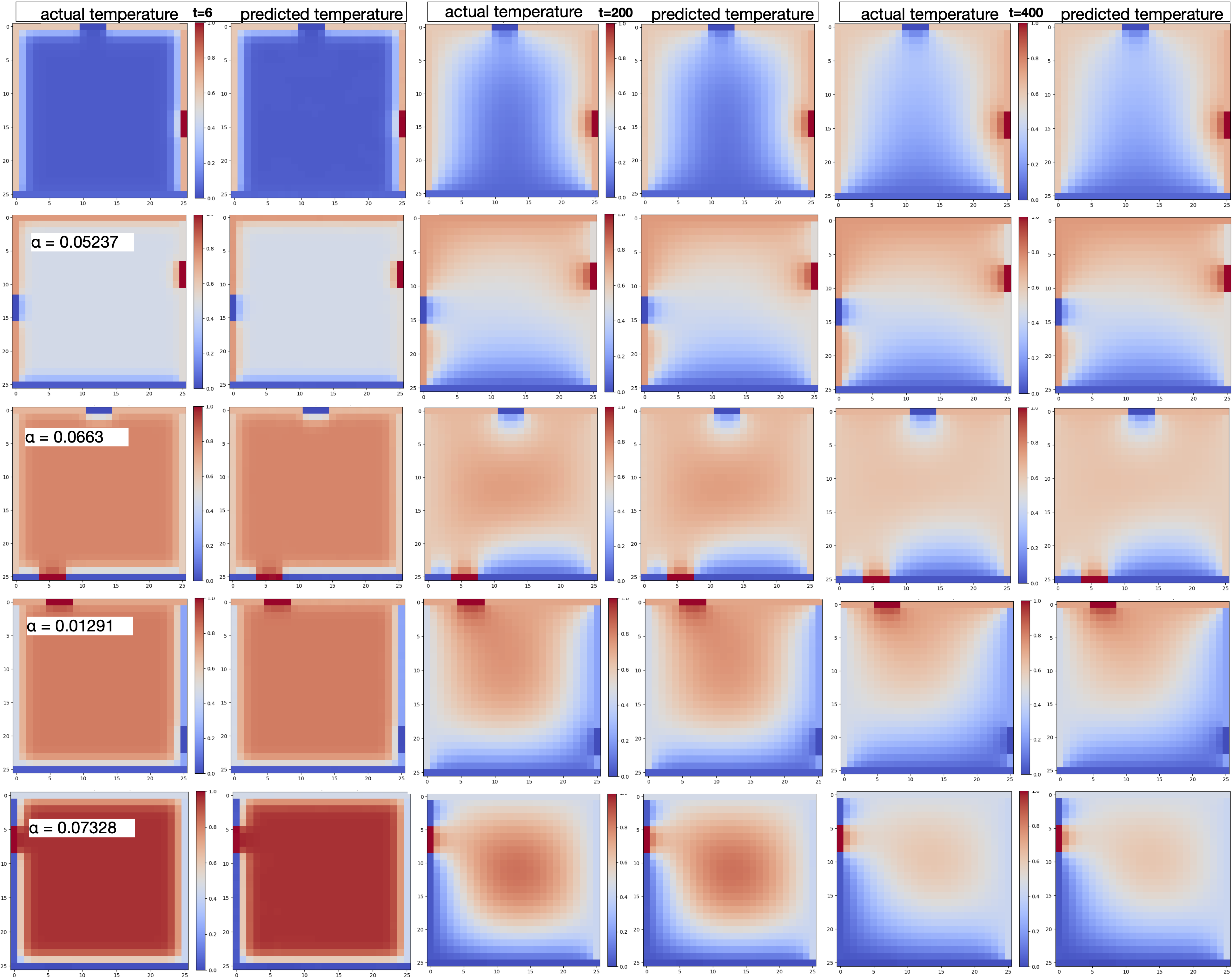}
        \caption{Same as Fig.~\ref{fig:Fames_Base_Block} but  for inference in autoregressive prediction mode in Challenge-2. }
    \label{fig:Challenge2_regressive}
\end{figure}

The model performance in the autoregressive inference mode is shown shown in Figure~\ref{fig:Challenge2_regressive} for a set of randomly cases from the hidden test set. The model is able to capture successfully the regions of increased local temperature variability associated with the randomly placed boundary segments.  

\vfill
\eject
\break

\section{Discussion, Limitations and Insights}
\label{section:Discussion}

Our aim in this work has been to offer a friendly introduction to the application of Transformer models in engineering physics. For this reason, we have discussed the key ingredients of Transformer NNs and drew parallels to equivalent mathematical concepts commonly used in Engineering. For the same reason, we have chosen to specifically focus on the 2D heat conduction problem as an example application because it is governed by a classical partial differential equation whose numerical solution is well understood . However, it is important to recognize that several optimizations and explorations have been intentionally left out of this introductory paper to be addressed in future work. Since the code is shared under an MIT license on GitHub\cite{mlxbeyond2024}, we invite others to explore some these directions as well. 

Some of the limitations to be addressed in future work include:

\begin{itemize}
	\item \textit{Speed Optimization:} While MLX provides significant speed improvements due to its use of unified memory, further optimizations could be explored, particularly in terms of memory management and parallelization on larger datasets.
    \item \textit{Distributed Parallel Computing:} MLX supports distributed computing on several GPUs over MPI (Message Passing Interface). We have shown that the model can be trained and applied using a small cluster of four Mac Studios with M2 Ultra processors. However, we have not yet attempted a thorough performance benchmark because, on one hand,  potential gains are tied to the details of the distributed training strategy and on the other, the distributed computing capabilities of MLX are still under development.  
	\item \textit{Exploration of Training Strategies:} There are various training strategies that could be further investigated, such as experimenting with different optimizers (e.g., AdamW, LAMB) and learning rate schedules. The effect of different batch sizes and gradient accumulation strategies on model performance could also be explored.
		\item \textit{Exploration of  Cross-Correlation in Loss}: While the present implementation already captures essential temporal correlations via the physics-informed loss, future work could  explore enforcing cross-correlation as part of the loss computation to complement the physics-informed loss component in capturing time evolution dynamics. This exploration could be particularly important for applications where the governing partial differential equations include stochasticity.   
	\item \textit{Minimal Model Configurations:} Although the current model configuration (number of layers, attention heads, etc.) has proven effective, a more systematic exploration of minimal configurations could help reduce model complexity while maintaining performance, particularly for deployment on lower-memory devices.
	\item \textit{Comparison to Other Neural Networks:} A thorough comparison of Transformers with other neural network architectures commonly used in physics simulations, such as convolutional neural networks (CNNs) or recurrent neural networks (RNNs), could provide deeper insights into the relative strengths and weaknesses of each approach.
	\item \textit{Scalability to Larger Systems:} Future work will involve testing the model’s ability to generalize to more complex boundary conditions, higher-dimensional PDEs, and larger grids. This includes assessing the model’s behavior when scaling up both spatial resolution and temporal duration.
\end{itemize}

\begin{figure}[h!!]
    \centering
    \includegraphics[width=1.0\textwidth]{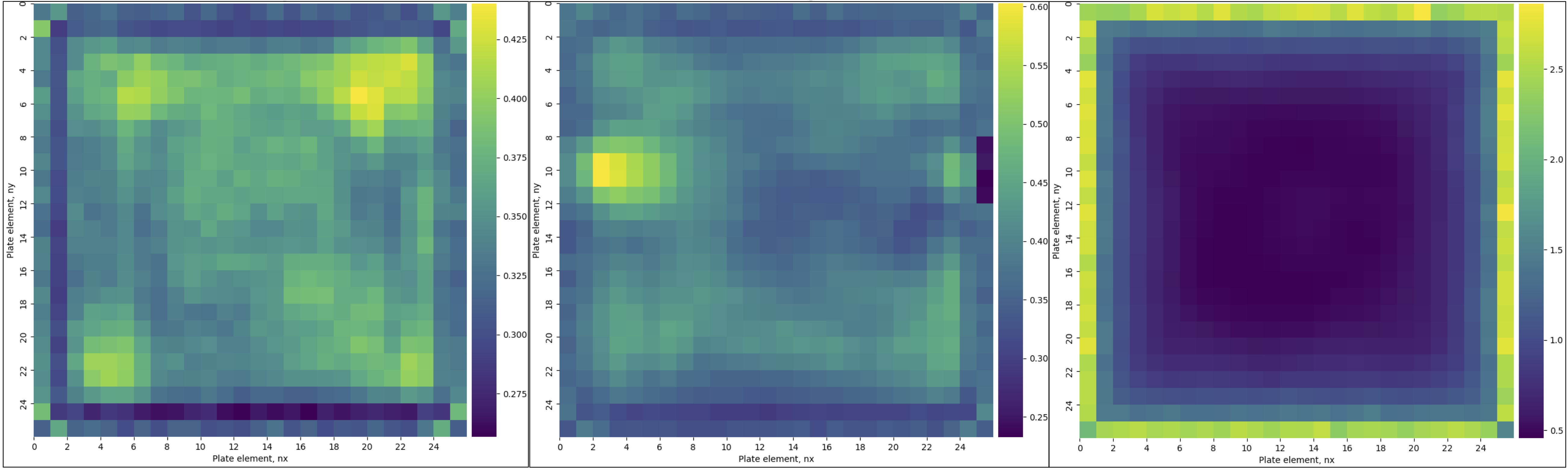}

    \caption{The average of the absolute weights for the final projection layer of the Transformer  model trained for block predictions. The leftmost panel corresponds to the base case, the middle one to Challenge-1 and the rightmost to Challenge-2.}
    \label{fig:insightsWeightsBlock}
\end{figure}

Additional insight in the workings of the Transformer  model can be gained by looking at the weights of the final projection layer for the Transformer  output at the end of training. In all three cases, the maximum absolute weight values (not shown here) occur on the Dirichlet boundaries, which is not surprising given that the Dirichlet boundary conditions  constitute the main distinguishing feature of each case. A more interesting view, however, is that of the \emph{average} of the absolute weight values which gives us a more holistic sense of where the model places emphasis. The average is computed over the latent dimension to create the 2D heatmap, shown in Figure~\ref{fig:insightsWeightsBlock}. The leftmost panel corresponds to for the base case, the middle panel to those for the Challenge-1 and the rightmost panel for Challenge-2. All there panels correspond to the model trained for block predictions. 

In the base case, on the average the model appears to place more emphasis on the internal nodes in the vicinity of the four domain corners where the effects of different Dirichlet temperatures merge to create regions with more pronounced gradients and variability. In the case of Challenge-1, the model continues to put emphasis in regions of the four domain corners, but places higher emphasis in the interior nodes adjacent to the boundary segments. This reflects the fact the interaction of the random Dirichlet values, which differ from one case to the next, with the segments create regions of variability that the model must be able to capture adequately. Thus, the model has been able to adapt to the localized variability caused by the introduction of the boundary segments. Finally. in the case of Challenge-2, where the two segments with temperatures T=1 and T=0 were randomly placed on separate sides of the plate, the model adapted by focusing on multiple perimetric rings of nodes adjacent to the edges. This adaptation likely enables the model to capture localized temperature variations induced by the random segments. The fact the placement of the segments can be anywhere on the perimeter of the domain has forced the model to place emphasis in concentric rings with weights that decrease from the periphery of the domain towards the center. These findings highlight the model’s sophisticated understanding of boundary conditions and its ability to generalize and adapt to new scenarios, underscoring the potential of Transformer  models in solving complex engineering physics problems.

\section{Conclusion and Broader Outlook}
\label{section:Conclusion}

In this work we aimed to offer a gentle introduction to using Transformers to solve Engineering problems, in particular problems that are governed by partial differential equations. The work is addressed to  engineers with working understanding of neural networks in general, but with no or limited exposure to Transformers. We have introduced the key concepts of Transformer models in their native context of Natural Language Processing, but where possible have drawn analogies between these features of Transformers and concepts that engineers are familiar with, such as the analogy of the self-attention mechanism to discrete Fourier transforms and of positional embeddings to wavelets. Since the primary mission in this work was to demystify Transformers and allow others to explore their use in Engineering, we have selected the problem of heat conduction in a 2D plate with Dirichlet boundary conditions, and various additional complications, as the playground where to exemplify the process of training a Transformer to predict the temporal evolution of a field variable (in this case the temperature), while capturing long-term temporal and spatial relations. In all of the considered configurations, we have used a completely hidden test set to show that the model achieved excellent performance both in block and autogressive inference mode. 

 With the potential of Transformers established, we next plan to explore the performance of this class of NNs in more challenging applications. In particular, we are working on implementing Transformer-augmented subgrid models for use with Large Eddy Simulations of turbulent fluid flows. The motivation is to leverage the inherent ability of transformers to capture long-term temporal and spatial relations, a trademark feature of turbulent flows. The task is markedly more challenging than the simple problem discussed herein, but the potential impact is significant and drives our efforts. We hope to report on our progress soon. 

 We hope this work will be useful to others who want to explore the use of Transformers in the engineering applications and towards this end, we have shared the code on GitHub under an MIT license\cite{mlxbeyond2024}. We will keep updating the code as our own experience with Transformers expands and future updates will be reflected on the GitHub distribution. 



\vfill\eject

\appendix
\section{The MLX Transformer Class}
\label{sec:appendixA}

The entire code used in this work is available under an MIT license on GitHub\cite{mlxbeyond2024}. Here, however, we provide the code defining the MLX Transformer class along with the model parameters used in tabular form. 

\lstset{
    language=Python,
    basicstyle=\ttfamily\scriptsize,  
    keywordstyle=\color{blue},
    commentstyle=\color{gray},
    stringstyle=\color{red},
    showstringspaces=false,
    frame=single,
    breaklines=true
}

\begin{lstlisting}
import mlx.nn as nn
import mlx.core as mx
import mlx.optimizers as optim

class HeatDiffusionModel(nn.Module):
    def __init__(self, ny, nx, seq_len, num_heads, num_encoder_layers,
                 mlp_dim, embed_dim, start_predicting_from, mask_type):
        super().__init__()
        self.seq_len = seq_len
        self.output_seq_len = seq_len
        self.ny = ny
        self.nx = nx
        self.input_dim = ny * nx
        self.embed_dim = embed_dim
        self.spatial_features = embed_dim // 2
        self._start_predicting_from = start_predicting_from
        self._mask_type = mask_type

        self.projection_spatial_enc = nn.Linear(
            ny * nx * self.spatial_features, self.embed_dim)

        self.positional_encoding_y = nn.SinusoidalPositionalEncoding(
            dims=self.spatial_features, max_freq=1, cos_first=False,
            scale=(1. / (np.sqrt(self.spatial_features // 2))), full_turns=False)
        self.positional_encoding_x = nn.SinusoidalPositionalEncoding(
            dims=self.spatial_features, max_freq=1, cos_first=False,
            scale=(1. / (np.sqrt(self.spatial_features // 2))), full_turns=False)
        self.positional_encoding_t = nn.SinusoidalPositionalEncoding(
            dims=self.embed_dim, max_freq=1,
            scale=(1. / (np.sqrt(self.embed_dim// 2))), full_turns=False)

        self.transformer_encoder = nn.TransformerEncoder(
            num_layers=num_encoder_layers, dims=embed_dim, num_heads=num_heads,
            mlp_dims=mlp_dim, checkpoint=False)

        self.output_projection = nn.Linear(embed_dim, ny * nx)

        self.diffusivity_embedding = nn.Linear(1, embed_dim)

        self.layer_normalizer = nn.LayerNorm(dims=embed_dim)

        if self._mask_type == 'causal':
            self.mask = self.create_src_causal_mask(self.seq_len)
        elif self._mask_type == 'block':
            self.mask = self.create_src_block_mask(self.seq_len)
        else:
            raise ValueError("Unsupported mask type")

    def create_src_block_mask(self, seq_len):
        mask = mx.full((seq_len, seq_len), -mx.inf, dtype=mx.float32)
        mask[:, :self._start_predicting_from] = 0
        return mask

    def create_src_causal_mask(self, seq_len):
        mask = mx.triu(-mx.inf * mx.ones((seq_len, seq_len)), k=0)
        mask[:, :self._start_predicting_from] = 0
        return mask

    def create_tgt_causal_mask(self, seq_len):
        mask = mx.triu(-mx.inf * mx.ones((seq_len, seq_len)), k=0)
        mask[:, :self._start_predicting_from] = 0
        return mask

    def __call__(self, src, alpha):
        batch_size, seq_len, _, _ = src.shape
        src_unflattened = src[:, :, :]
        src_expanded = mx.expand_dims(src_unflattened, -1)
        pos_enc_ny, pos_enc_nx = self.spatial_positional_encoding(self.ny, self.nx)
        src_pos_enc_y = src_expanded + pos_enc_ny
        src_pos_enc = src_pos_enc_y + pos_enc_nx
        src_pos_enc_flattened = src_pos_enc[:, :, :, :].reshape(-1, seq_len,
                                                                self.ny * self.nx * self.spatial_features)
        src_projected = self.projection_spatial_enc(src_pos_enc_flattened)

        temporal_enc = self.temporal_positional_encoding(seq_len, batch_size)

        src_encoded = src_projected + temporal_enc

        alpha_reshaped = alpha.reshape(-1, 1)
        alpha_embed = self.diffusivity_embedding(alpha_reshaped)

        alpha_embed_expanded = mx.expand_dims(alpha_embed, axis=1)
        alpha_embed_expanded = mx.broadcast_to(alpha_embed_expanded,
                                    (batch_size, seq_len, self.embed_dim))

        src_encoded += alpha_embed_expanded

        encoded = self.transformer_encoder(src_encoded, mask=self.mask)

        normalized = self.layer_normalizer(encoded)

        output = self.output_projection(normalized)

        output = output.reshape(batch_size, self.output_seq_len, self.ny, self.nx)

        return output

    def spatial_positional_encoding(self, ny, nx):
        nx_encoding = mx.expand_dims(mx.expand_dims(mx.expand_dims(
            self.positional_encoding_x(mx.arange(self.nx)), 0), 0), 1)
        ny_encoding = mx.expand_dims(mx.expand_dims(mx.expand_dims(
            self.positional_encoding_y(mx.arange(self.ny)), 0), 0), 3)
        return ny_encoding, nx_encoding

    def temporal_positional_encoding(self, seq_len, batch_size):
        temporal_encoding = mx.expand_dims(self.positional_encoding_t(mx.arange(self.seq_len)), axis=0)
        return temporal_encoding

\end{lstlisting}

\begin{table}[htbp]
\centering
\begin{tabular}{|>{\raggedright}m{4.5cm}|>{\raggedright}m{4.5cm}|>{\centering\arraybackslash}m{1.2cm}|>{\centering\arraybackslash}m{1.2cm}|}
\hline
\textbf{Parameter Name} & \textbf{Role/Explanation} & \textbf{Base, Ch-1} & \textbf{Ch-2} \\
\hline
\texttt{start\_predicting\_from} & Defines the time step to start making predictions & 5 & 5 \\
\hline
\texttt{batch\_size} & The number of samples processed before model parameter updates & 4 & 4 \\
\hline
\texttt{epochs} & Number of times the model goes through the entire dataset & 100 & 300 \\
\hline
\texttt{seq\_len} & Length of the input sequence for each sample & 401 & 401 \\
\hline
\texttt{num\_heads} & Number of attention heads in the multi-head attention mechanism & 16 & 16 \\
\hline
\texttt{num\_encoder\_layers} & Number of encoder layers in the transformer & 12 & 24 \\
\hline
\texttt{mlp\_dim} & Size of the hidden layer in the feedforward network of the transformer block & 256 & 256 \\
\hline
\texttt{embed\_dim} & Dimensionality of the input and output embeddings & 512 & 512 \\
\hline
\texttt{mask\_type} & Type of attention masking applied to control which parts of the sequence the model should focus on & block or causal & block or causal \\
\hline
\end{tabular}
\caption{Comparison of Parameters for Base, Challenge-1 (Ch-1) and Challenge-2 (Ch-2) Cases.}
\end{table}





\end{document}